%% file: DESY-06-144.tex
\documentclass[zpreprint,zbstdefault]{./zeus_paper}
\usepackage{subfigure}
\usepackage[english]{babel}

\input ./zeus_def.tex
\input ./def.tex

\def\citeCTD{{\cite{%
nim:a279:290,*npps:b32:181,*nim:a338:254%
}}\xspace}
\def\citeCAL{{\cite{%
nim:a309:77,*nim:a309:101,*nim:a321:356,*nim:a336:23%
}}\xspace}
\begin{document}
%------------------------------------------------------------------------------
%       Title sheet
%------------------------------------------------------------------------------
%\include{DESY-06-144-tit}
%%%%%%%%%%%%%%%%%%%%%%%%%%%%%%%%%%%%%%%%%%%%%%%%%%%%%%%%%%%%%%%%%%%%%%%%%%%%%%%
\title{
Search for stop production in $R$-parity-violating supersymmetry at HERA
}                                                       
                    
\author{ZEUS Collaboration}
\prepnum{DESY 06-144}
\date{\today}

\abstract{  A  search  for  stop  production  in  $R$-parity-violating
supersymmetry  has been  performed in  $e^{+}p$ interactions  with the
ZEUS detector at HERA, using an integrated luminosity of 65~pb$^{-1}$.
At HERA, the  $R$-parity-violating coupling $\lambda'$ allows resonant
squark  production, $e^+d\rightarrow\tilde{q}$. Since  the lowest-mass
squark  state  in  most   supersymmetry  models  is  the  light  stop,
$\tilde{t}$,  this search concentrated  on production  of $\tilde{t}$,
followed  either by  a direct  $R$-parity-violating decay,  or  by the
gauge decay to $b\tilde{\chi}^+_{1}$.  No evidence for stop production
was found and limits were set on $\lambda'_{131}$ as a function of the
stop  mass in  the framework  of the  Minimal  Supersymmetric Standard
Model. The results have also  been interpreted in terms of constraints
on the parameters of the minimal Supergravity model.}

\makezeustitle
\def\3{\ss}                                                                                        
%\newcommand{\address}{ }                                                                           
%\renewcommand{\author}{ }                                                                          
%\pagenumbering{Roman}                                                                              
                                    % this "%"s are for cosmetics only                             
%\begin{document}                                                                                   
                                                   %                                               
\begin{center}                                                                                     
{                      \Large  The ZEUS Collaboration              }                               
\end{center}                                                                                       
  S.~Chekanov$^{   1}$,                                                                            
  M.~Derrick,                                                                                      
  S.~Magill,                                                                                       
  S.~Miglioranzi$^{   2}$,                                                                         
  B.~Musgrave,                                                                                     
  D.~Nicholass$^{   2}$,                                                                           
  \mbox{J.~Repond},                                                                                
  R.~Yoshida\\                                                                                     
 {\it Argonne National Laboratory, Argonne, Illinois 60439-4815}, USA~$^{n}$                       
\par \filbreak                                                                                     
  M.C.K.~Mattingly \\                                                                              
 {\it Andrews University, Berrien Springs, Michigan 49104-0380}, USA                               
\par \filbreak                                                                                     
  N.~Pavel~$^{\dagger}$, A.G.~Yag\"ues Molina \\                                                   
  {\it Institut f\"ur Physik der Humboldt-Universit\"at zu Berlin,                                 
           Berlin, Germany}
% ...
\newpage                                                                                           
$^{\    1}$ supported by DESY, Germany \\                                                          
$^{\    2}$ also affiliated with University College London, UK \\                                  
% ...
$^{\dagger}$ deceased \\                                                                           
%                                                                                                  
% \par         % if index listing & table fit to 1 page, put gap here                              
\newpage   % alternatively: go to newpage, if page is too small                                    
                                                           %                                       
% \institute_references_start    % do not touch or move this line !                                
                                                           %                                       
\begin{tabular}[h]{rp{14cm}}                                                                       
$^{a}$ &  supported by the Natural Sciences and Engineering Research Council of Canada (NSERC) \\  
$^{b}$ &  supported by the German Federal Ministry for Education and Research (BMBF), under        
          contract numbers HZ1GUA 2, HZ1GUB 0, HZ1PDA 5, HZ1VFA 5\\                                
$^{c}$ &  supported in part by the MINERVA Gesellschaft f\"ur Forschung GmbH, the Israel Science   
          Foundation (grant no. 293/02-11.2) and the U.S.-Israel Binational Science Foundation \\  
$^{d}$ &  supported by the German-Israeli Foundation and the Israel Science Foundation\\           
$^{e}$ &  supported by the Italian National Institute for Nuclear Physics (INFN) \\                
$^{f}$ &  supported by the Japanese Ministry of Education, Culture, Sports, Science and Technology 
          (MEXT) and its grants for Scientific Research\\                                          
$^{g}$ &  supported by the Korean Ministry of Education and Korea Science and Engineering          
          Foundation\\                                                                             
$^{h}$ &  supported by the Netherlands Foundation for Research on Matter (FOM)\\                   
$^{i}$ &  supported by the Polish State Committee for Scientific Research, grant no.               
          620/E-77/SPB/DESY/P-03/DZ 117/2003-2005 and grant no. 1P03B07427/2004-2006\\             
$^{j}$ &  partially supported by the German Federal Ministry for Education and Research (BMBF)\\   
$^{k}$ &  supported by RF Presidential grant N 1685.2003.2 for the leading scientific schools and  
          by the Russian Ministry of Education and Science through its grant for Scientific        
          Research on High Energy Physics\\                                                        
$^{l}$ &  supported by the Spanish Ministry of Education and Science through funds provided by     
          CICYT\\                                                                                  
$^{m}$ &  supported by the Particle Physics and Astronomy Research Council, UK\\                   
$^{n}$ &  supported by the US Department of Energy\\                                               
$^{o}$ &  supported by the US National Science Foundation\\                                        
$^{p}$ &  supported by the Polish Ministry of Scientific Research and Information Technology,      
          grant no. 112/E-356/SPUB/DESY/P-03/DZ 116/2003-2005 and 1 P03B 065 27\\                  
$^{q}$ &  supported by FNRS and its associated funds (IISN and FRIA) and by an Inter-University    
          Attraction Poles Programme subsidised by the Belgian Federal Science Policy Office\\     
$^{r}$ &  supported by the Malaysian Ministry of Science, Technology and                           
Innovation/Akademi Sains Malaysia grant SAGA 66-02-03-0048\\                                       
\end{tabular}                                                                                      
                                                           %                                       
% \institute_references_end     % do not touch or move this line !                                 
                                                           %                                       
%\end{document}                                                                                     
%%%%%%%%%%%%%%%%%%%%%%%%%%%%%%%%%%%%%%%%%%%%%%%%%%%%%%%%%%%%%%%%%%%%%%%%%%%%%%%
%------------------------------------------------------------------------------
%       Text
%------------------------------------------------------------------------------
%\include{DESY-06-144-txt}
%%%%%%%%%%%%%%%%%%%%%%%%%%%%%%%%%%%%%%%%%%%%%%%%%%%%%%%%%%%%%%%%%%%%%%%%%%%%%%%
\newpage
\pagenumbering{arabic} 
\pagestyle{plain}

%%%%%%%%%%%%%%%%%%%%%%%%%%%%%%%%%%%%%%%%%%%%%%%%%%%%%%%%%%%%%%%%%%%%%%
% ----------------------------------------------------------------------------
%       Introduction
% ----------------------------------------------------------------------------
%%%%%%%%%%%%%%%%%%%%%%%%%%%%%%%%%%%%%%%%%%%%%%%%%%%%%%%%%%%%%%%%%%%%%%

\section{Introduction}
\label{sec-int}

Many extensions of  the Standard Model (SM) require  a new fundamental
symmetry  between bosons and  fermions, known  as \emph{supersymmetry}
(SUSY)~\cite{SUSY}.  This  symmetry,  hypothesizing the  existence  of
supersymmetric partners  of the SM particles,  with similar properties
but with spin changed by one half, controls the divergent higher-order
loop corrections  to the  Higgs-boson mass. Despite  numerous searches
for such new particles, no evidence has been observed, indicating that
supersymmetry,  if it exists,  becomes manifest  at scales  beyond the
present experimental  limit. Since  SUSY involves so  many parameters,
different experimental techniques complement each other.

One important  quantum number  in supersymmetry models  is $R$-parity
($R_{p}$). Its  conservation ensures  the conservation of  both lepton
and baryon  number. Most of  the searches performed at  colliders drew
conclusions under the assumption of $R_{p}$ conservation. Nevertheless,
the  most  general  supersymmetric  extension  of  the  SM  Lagrangian
contains  terms  which  violate  $R_{p}$  and  some  of  the  possible
$R_{p}$-violating ($\rpv$)  scenarios are compatible  with the present
experimental constraints.

One of  the most interesting  consequences of $\rpv$ scenarios  is the
possibility  of  producing   single  SUSY  particles  (sparticles)  at
colliders. Electron-proton  collisions at HERA are well  suited to the
search  for squarks,  the  scalar supersymmetric  partners of  quarks,
since such  states can be produced  by an appropriate  coupling of the
incoming lepton and a quark in the proton.

In most of the SUSY scenarios, the squarks of the third generation are
the lightest; the present analysis is aimed of searching for the stop,
$\sTop$, the  supersymmetric partner  of the top  quark. At  HERA, the
stop can be produced resonantly via $e^{+}d \to \sTop$, up to the $ep$
centre-of-mass energy  $\sqrt{s} \simeq 320 \gev$. The  stop decay can
lead to distinctive topologies with a high-energy positron or neutrino
and  hadronic jets,  which can  be efficiently  separated from  the SM
background.

Direct  searches for stop  production have  been already  performed at
HERA\cite{H1:SUSY},     LEP\cite{opal:stop,aleph:stop,l3:stop}     and
Tevatron\cite{cdf:stop}.

%%%%%%%%%%%%%%%%%%%%%%%%%%%%%%%%%%%%%%%%%%%%%%%%%%%%%%%%%%%%%%%%%%%%%%
% ----------------------------------------------------------------------------
%       Phenomenology
% ----------------------------------------------------------------------------
%%%%%%%%%%%%%%%%%%%%%%%%%%%%%%%%%%%%%%%%%%%%%%%%%%%%%%%%%%%%%%%%%%%%%%

\section{Stop phenomenology}
\label{sec-pheno}

The $R$-parity, defined as $R_{p}=(-1)^{3B+L+2S}$, is a multiplicative
quantum number which is 1  for particles and $-1$ for sparticles ($B$,
$L$   and  $S$  denote   baryon  number,   lepton  number   and  spin,
respectively). $R$-parity conservation would imply that supersymmetric
particles  are always  pair produced  and the  lightest supersymmetric
particle (LSP) is stable, a good candidate for cold dark matter.

If $R$-parity is violated, it  is possible to create single sparticles
that decay to  SM particles~\cite{butt}. The $\rpv$ terms  in the SUSY
superpotential are given by:

\begin{equation}  W_{ \rpvdue} =
\lambda_{\emph{ijk}} L_{\emph{i}} L_{\emph{j}} \overline{E}_{\emph{k}}
+        \lambda'_{\emph{ijk}}        L_{\emph{i}}        Q_{\emph{j}}
\overline{D}_{\emph{k}}    +    \lambda''_{\emph{ijk}}    U_{\emph{i}}
\overline{D}_{\emph{j}} \overline{D}_{\emph{k}}, 
\end{equation} 

where  the   subscripts  \emph{i,j,k}  are   the  generation  indices,
$L_{\emph{i}}$   denotes  the   $SU(2)$  doublet   lepton  superfield,
$E_{\emph{i}}$ the  $SU(2)$ singlet lepton  superfield, $Q_{\emph{i}}$
the   $SU(2)$  doublet   quark  superfield   and   $D_{\emph{i}}$  and
$U_{\emph{i}}$   the   $SU(2)$  singlet   down-   and  up-type   quark
superfields.       The       dimensionless      Yukawa       couplings
$\lambda_{\emph{ijk}}$,                        $\lambda'_{\emph{ijk}}$,
$\lambda''_{\emph{ijk}}$ are free parameters of the model.

In  the case of  stop production  in $ep$  collisions, the  only terms
involved are those parameterized  by the Yukawa coupling $\lprim$. The
partners  of  the  left-   and  right-handed  top,  $\tilde{t}_L$  and
$\tilde{t}_R$, can mix together in two mass eigenstates, $\tilde{t}_1$
and $\tilde{t}_2$, which, because of  the large top mass, are usually
strongly  non-degenerate. Due  to the  chiral properties  of  the SUSY
superpotential (Eq.~1) only the $\tilde{t}_L$ state contributes to the
stop  production   cross  section.  In  the   regime  of  narrow-width
approximation  (NWA), the  production  cross section  for the  lighter
state $\tilde{t}_{1}$ is hence:

\begin{equation}
\sigma(e^{+}p \to \tilde{t}_{1})= \frac{\pi}{4s}~(\cos{\theta_{\tilde{t}}} \cdot \lprim )^{2} ~ d(M_{\stone}^{2}/s,M_{\stone}^{2}),
\end{equation}

where $d(x, Q^{2})$\footnote{ The  variables $x$, $Q^2$ and $y$, which
  is  used later  on in  the  paper, are  the three  Lorentz-invariant
  quantities characterizing  the DIS processes. $Q^2$  is the negative
of the four-momentum-transfer  squared,  $x$  the  fraction of  the  proton
  momentum carried by  the struck quark and $y$  the inelasticity.} is
the   parton  density   of   the   $d$  quark   in   the  proton   and
$\theta_{\tilde{t}}$ is  the mixing angle  between $\tilde{t}_{1}$ and
$\tilde{t}_{2}$. In  this paper only the  lighter stop, $\tilde{t}_1$,
denoted   hereafter  as   $\tilde{t}$,  has   been   considered  since
contributions from $\tilde{t}_2$ are negligible for all the considered
scenarios.
%%%%% Start correction dded by Lorenzo
The effects of  the initial state photon radiation  decreases the stop
production cross section of $\sim 5\%$ ($20\%$) for a stop mass of 150
$(280) \gev$   and   have   been   taken  into   account   using   the
Weisz\"acker-Williams approach\cite{isr}. The NLO QCD corrections have
been also  included\cite{nlo}, they increase  the LO cross  section of
$\sim 20-25\%$.
%%%%% Finish correction added by Lorenzo

The   $\tilde{t}$   decays   considered   in  this   study   are   the
$R_{p}$-violating  channel   $\tilde{t}\rightarrow  e^+  d$   and  the
$R_{p}$-conserving decay  $\tilde{t} \rightarrow \tilde{\chi}_{1}^{+}
b$, where $\tilde{\chi}_{1}^{+}$ is the lightest chargino\footnote{The
superpartners of the  charged $SU(2)$ gauge bosons and  of the charged
Higgs bosons  mix together in two mass  eigenstates named charginos.}.
These two channels provide  a sufficiently large total branching ratio
over  all  of  the   considered  SUSY  parameter  space.  The  channel
$\tilde{t}\rightarrow\tilde{\chi}^{0}_{1}           t$,          where
$\tilde{\chi}^{0}_{1}$   is   the   lightest   neutralino\footnote{The
superpartners of the  neutral $SU(2)$ gauge bosons and  of the neutral
Higgs   bosons   mix  together   in   four   mass  eigenstates   named
neutralinos.}, contributes  only at the highest stop  masses, and even
then it is  below 10\%. Branching ratios involving  a heavier chargino
or neutralino are small in most of the considered parameter space.

The    considered   decays,   including    the   cascade    from   the
$\tilde{\chi}^{+}_{1}$,  are  illustrated in  Fig.~1.  In the  present
paper only the hadronic $W$  decays are considered, hence final states
involving one  positron with  one jet ($e$-J),  or more than  one jets
($e$-MJ), and one neutrino and multiple-jets ($\nu$-MJ) are studied.

The results have been interpreted in the context of two different SUSY
scenarios: the  Minimal Supersymmetric  Standard Model (MSSM)  and the
minimal Supergravity model (mSUGRA).

In the unconstrained $\rpv$  MSSM~\cite{APV}, the branching ratios for
stop decay  as well  as the masses  of the neutralinos,  charginos and
gluinos\footnote{Supersymmetric partners of gluons.} are determined by
the following  MSSM parameters: the  mass term $\mu$, which  mixes the
Higgs superfields; the soft  SUSY-breaking parameters $M_1$, $M_2$ and
$M_3$  for the  $U(1)$,  $SU(2)$ and  $SU(3)$ gauginos,  respectively;
$\tan{\beta}$, the ratio  of the vacuum expectation values  of the two
neutral  scalar Higgs  fields; and  the $\rpv$  Yukawa  couplings. The
search for the stop was performed  in the mass range $100 - 280 \gev$.
In order  to reduce the  number of free parameters,  the assumptions
listed below were made:

\begin{itemize}
\item no mixing between $\tilde{t}_{L}$ and  $\tilde{t}_{R}$  ($\theta_{\tilde{t}}=0$);
\item only the Yukawa coupling $\lambda'_{131}$ was assumed to be 
      non-zero;
\item SUSY scenarios in which the lightest neutralino $\tilde{\chi}^{0}_{1}$
      is not the LSP or is lighter than $30 \gev$, already excluded  by LEP 
      result~\cite{epj:c19:397},  were not considered;
\item the  GUT relations~\cite{polonsky}  for the  gaugino mass
      terms $M_1$, $M_2$ and $M_3$:
\begin{eqnarray*}
M_{1} &=& \frac{5}{3} \mbox{tan}^{2} \theta_{W} \cdot M_{2}, \\
M_{3} &=& \frac{\alpha_{S}}{\alpha_{\rm EM}} \mbox{sin}^{2} \theta_{W} \cdot M_{2}
\end{eqnarray*}
      were assumed.
      As a consequence, the  gluino is always heavier than  $\tilde{t}$, 
      so  that the decay  $\tilde{t} \to  t $~$ \tilde{g}$ is kinematically forbidden;
\item all the other  sfermions  (apart from the  lighter stop)  were assumed  
      to have large masses that were fixed at 1 TeV.
\end{itemize}

The  total branching  ratio due  to the  considered decay  channels is
$\gtrsim 70  \%$ for the  range of parameters  $|\mu| < 300  \gev$ and
$M_{2}$  between  100 and  $300  \gev$  that  was considered  in  this
analysis.

Figure~2 shows the branching  ratios for four representative points in
the   $\mu$-$M_{2}$  plane   involving  very   different   masses  for
$\tilde{\chi}^{0}_{1}$    and     $\tilde{\chi}^{+}_{1}$    and    for
$\lambda'_{131}$ values  equal to the limit given  in Section~6.2. The
shaded band  indicates the  sum of the  three considered  channels for
$2<\tan\beta<50$. The  total branching  ratio and the  contribution of
the three  channels are  shown for the  value $\tan\beta=6$.  When the
$\tilde{\chi}^{+}_{1}$ mass  is larger than the stop  mass, the $\rpv$
decay ($e$-J) is the dominant channel; in the other cases the $\nu$-MJ
channel is generally the most relevant.

In the mSUGRA~\cite{msugra}, the  number of free parameters is further
reduced by  assuming two universal  mass parameters at the  GUT scale,
$m_{0}$ and $m_{1/2}$, for all the sfermions and for all the gauginos,
respectively.   Radiative  corrections  are   assumed  to   drive  the
electroweak   symmetry  breaking   (REWSB),  leading   to  consistency
relations that  allow the  complete model to  be fixed, based  only on
$m_0$,  $m_{1/2}$, the sign  of $\mu$,  $\tan{\beta}$, and  the common
trilinear  coupling $A_0$.  In the  range of  parameters used  in this
analysis,  $m_{1/2} <  180 \gev$  and $m_{0}  < 300  \gev$,  the total
branching ratio of  the considered channels is in  the range $0.4-0.8$
for  $m_{0} < 200  \gev$. For  larger values  of $m_{0}$  it decreases
rapidly, since  decay channels  involving the heavier  chargino, which
are not considered in this study, become important.

%%%%%%%%%%%%%%%%%%%%%%%%%%%%%%%%%%%%%%%%%%%%%%%%%%%%%%%%%%%%%%%%%%%%%%
% ----------------------------------------------------------------------------
%       Experimental set-up
% ----------------------------------------------------------------------------
%%%%%%%%%%%%%%%%%%%%%%%%%%%%%%%%%%%%%%%%%%%%%%%%%%%%%%%%%%%%%%%%%%%%%%

\section{Data sample and experimental set-up}

% Here you might want to describe your data samples and the detector. In
% particular, you may use some of the text fragments provided as \LaTeXZ
% macros.

The data used in this  analysis were collected in the years 1999-2000.
The total integrated luminosity was $65.1\pm1.5$ pb$^{-1}$ of $e^{+}p$
collisions.   The   proton   and   positron  energies   were   $E^{\rm
beam}_{p}=920\gev$  and  $E^{\rm  beam}_{e}=27.5 \gev$,  respectively,
leading to a centre-of-mass energy of $318 \gev$.

\Zdetdesc

\Zctddesc\ZcoosysfnBeta

\Zcaldesc

%\Zlumidesc

The luminosity was measured using the bremsstrahlung process $ep \to e
p \gamma$.  The resulting small-angle energetic  photons were measured
by  the  luminosity  monitor~\cite{Desy-92-066},  a  lead-scintillator
calorimeter placed in the HERA tunnel at $Z = -107$~m.

%%%%%%%%%%%%%%%%%%%%%%%%%%%%%%%%%%%%%%%%%%%%%%%%%%%%%%%%%%%%%%%%%%%%%%
% ----------------------------------------------------------------------------
%       Simulation
% ----------------------------------------------------------------------------
%%%%%%%%%%%%%%%%%%%%%%%%%%%%%%%%%%%%%%%%%%%%%%%%%%%%%%%%%%%%%%%%%%%%%%

\section{Monte Carlo simulation}

The signal  processes were simulated  with the Monte Carlo  (MC) event
generator  {\sc Susygen}~3~\cite{susygen}.  It uses  the  exact matrix
element for the production and  for the decays of sparticles, includes
initial- and  final-state radiative  corrections and is  interfaced to
{\sc  Pythia}~6.2~\cite{pythia}  for the  hadronization  of the  final
state. The program {\sc  Suspect}~2.1~\cite{suspect} was used to solve
the  REWSB consistency  relations  that determine  the sparticle  mass
spectrum at the electroweak scale in the mSUGRA model.

The dominant  background process to  the $e$-J and $e$-MJ  channels is
neutral  current deep  inelastic scattering  (NC DIS).  For  the $e$-J
channel, the background is due  to $2 \to 2$ scatters between high-$x$
quarks and the positron. Backgrounds to the $e$-MJ channel occur in NC
DIS events  where multi-jet final states result  from higher-order QCD
effects.

The  primary background  to the  $\nu$-MJ channel  comes  from charged
current deep inelastic scattering (CC DIS) with multiple jets from QCD
radiation.  An additional  background source  involves photoproduction
events  for which  the measured  transverse momentum  is large  due to
energy mismeasurement.

The NC  and CC  events were simulated  using the  {\sc Heracles}~4.6.1
\cite{heracles}  program   with  the  {\sc  Djangoh}~1.1~\cite{django}
interface    to   the    hadronization   program    and    using   the
CTEQ5D~\cite{cteq5}  set of parton  distribution functions  (PDFs). In
{\sc   Heracles},  corrections   for  initial-state   and  final-state
electroweak   radiation,  vertex   and  propagator   corrections,  and
two-boson  exchange  are included.  The  colour-dipole  model of  {\sc
Ariadne}~4.10~\cite{ariadne}   was   used   to  simulate   the   order
$\alpha_{S}$ plus leading-logarithmic  corrections to the quark-parton
model. The  MEPS model of  {\sc Lepto}~6.5~\cite{lepto} was used  as a
systematic  check. Both  programs use  the Lund  string model  of {\sc
Jetset}~7.4\cite{jetset} for the hadronization.

Photoproduction    events    were    simulated    using    the    {\sc
Herwig}~6.100~\cite{herwig}  generator, using  the CTEQ4L~\cite{cteq4}
proton PDFs. Both direct and resolved photoproduction were considered.
In the direct case, all of  the photon energy participates in the hard
scattering, whereas, for the resolved  process, only a fraction of the
photon  energy, associated with  a parton  constituent of  the photon,
participates  in  the  hard  subprocess.  For the  simulation  of  the
resolved subprocess, the GRV-G~\cite{grvg} photon PDFs were used.

%%%%%%%%%%%%%%%%%%%%%%%%%%%%%%%%%%%%%%%%%%%%%%%%%%%%%%%%%%%%%%%%%%%%%%
% ----------------------------------------------------------------------------
%       Event Selection
% ----------------------------------------------------------------------------
%%%%%%%%%%%%%%%%%%%%%%%%%%%%%%%%%%%%%%%%%%%%%%%%%%%%%%%%%%%%%%%%%%%%%%

\section{Event selection}
\label{sec-sel}

The signal  events are  characterized by a  high-energy lepton  in the
final state. In  the case of a positron in the  final state ($e$-J and
$e$-MJ  channels),  the trigger  selection  was  based  on a  standard
neutral current trigger, which  required a scattered positron, as used
in searches  for resonance states decaying  to $eq$~\cite{ZEUS:LQ} and
in ZEUS NC DIS studies~\cite{ZEUS:NC}. For the neutrino case ($\nu$-MJ
channel), a  trigger selection based on  a missing-$P_{T}$ requirement
and already  employed in the  ZEUS CC DIS  analysis~\cite{ZEUS:CC} was
used.
 
The   offline   signal-search   procedure   was   performed   in   two
steps~\cite{thesis}. Initially,  a preselection was  applied to select
NC or  CC events. Finally,  more restrictive selections,  optimized to
get the best limits in the case of no signal, were applied.

%%%%%%%%%%%%%%%%%%%%%%%%%%%%%%%%%%%%%%%%%%%%%%%%%%%%%%%%%%%%%%%%%%%%%%

\subsection{Preselection for $e^{+}$ final states}

The following conditions, some of  which were also used at the trigger
level  with a lower  threshold, were  designed to  select a  sample of
high-$Q^2$ NC events:
\begin{itemize}
\item  $Z$-coordinate  of  the  event  vertex  compatible  with  an  $ep$
  interaction, $|Z_{\rm vtx}|<50$ cm;
\item a high-energy positron reconstructed from calorimeter
  and tracking information~\cite{EM}. A positron energy $E_e > 8 \gev$
  was required. This cut was increased to $P_{T,e}>20 \gev$ ($P_{T,e}$
  is  the   transverse  momentum  of   the  positron  measured   by  the
  calorimeter) for very forward positrons ($\theta_e < 0.3$, where $\theta_e$ is the
  positron polar angle) which are outside the acceptance of the CTD;
\item $45< E-P_{Z} <  70 \gev$, where $E$ and $P_{Z}$ are the total 
  energy and the $Z$-component of the total momentum of the final state.
  For  NC  DIS  events,  where  only particles  in  the  very  forward
  direction   escape  detection, $E-P_Z \sim  2E^{\mbox{\scriptsize
  beam}}_e = 55 \gev$;
\item   $Q^2_{\mbox{\scriptsize  DA}}   >  1000 \gev{^2}$   and  $0.2
  <y_{\mbox{\scriptsize   DA}}<0.98$,   where  $Q^2_{\mbox{\scriptsize
      DA}}$  and   $y_{\mbox{\scriptsize  DA}}$  are   the  DIS
  kinematic variables reconstructed using the double angle method~\cite{DA}.  
  The above conditions were imposed in  order to restrict the search 
  to  a region where the signal is enhanced with respect  to NC DIS 
  and the reconstruction of the kinematic variables is reliable;
\item $M_{eX} > 100 \gev$, where  $M_{eX}$ is the invariant mass of the
  positron  and  the hadronic  system  evaluated  using the  following
  relation   that  exploits the conservation of  momentum and $E-P_{Z}$:
\begin{equation}
M^2_{eX}=2 E^{\mbox{\scriptsize beam}}_e \sum_i (E+P_Z)_i.
\end{equation}
  The  sum runs  over  the  final-state positron  and  all other  energy
  deposits with  a polar angle  $>0.1$, to exclude  contributions from
  the proton remnant. 
\end{itemize}

The bias and resolution of the reconstructed mass were evaluated using
the signal MC. On average,  the mass was slightly overestimated at low
masses  ($3\%$ at $100  \gev$), while  the agreement  improved towards
high masses ($< 1\%$ above  $150 \gev$). The resolution varied between
$5\%$  and $1.5\%$  in the  mass  range $100  - 280  \gev$. After  the
preselection cuts, $2368$ events  remained, in good agreement with the
expectation  of the  SM MC  of $2430^{+90}_{-252}$  events,  where the
error  is  dominated  by  the systematic  uncertainties  described  in
Section~\ref{sec-syst}. The  SM prediction is dominated by  the NC DIS
contribution.

Figure~3     shows     the     distributions     of     $P_{T,e}$,
$y_{\mbox{\scriptsize  DA}}$, $\log_{10}(Q^2_{\mbox{\scriptsize DA}})$
and  $P_{T,\mbox{\scriptsize  antipar}}/P_{T,\mbox{\scriptsize  par}}$
for data and  MC; reasonable agreement is seen  for all variables. The
quantities $P_{T,\mbox{\scriptsize  par}}$ and $P_{T,\mbox{\scriptsize
antipar}}$  are  the  parallel  and  antiparallel  components  of  the
hadronic  transverse momentum ($\vec{P}_{T,  \mbox{\scriptsize had}}$)
defined as:
\begin{eqnarray*}
%P_{T,\mbox{\scriptsize par}}     &=&   \sum_{i} \vec{P}_{T,i} \cdot \vec{n}_{P_{T,\mbox{\scriptsize had}}} \quad \mbox{for} \quad  \vec{P}_{T,i} \cdot \vec{n}_{P_{T,\mbox{\scriptsize had}}} > 0,\\
P_{T,\mbox{\scriptsize par}} &=& \frac{1}{|\vec{P}_{T,\mbox{\scriptsize had}}|} \cdot \sum_{\vec{P}_{T,i} \cdot \vec{P}_{T,\mbox{\tiny had}} > 0} \vec{P}_{T,i} \cdot \vec{P}_{T,\mbox{\scriptsize had}}, \\
%P_{T,\mbox{\scriptsize antipar}} &=& - \sum_{i} \vec{P}_{T,i} \cdot \vec{n}_{P_{T,\mbox{\scriptsize had}}} \quad \mbox{for} \quad \vec{P}_{T,i} \cdot \vec{n}_{P_{T,\mbox{\scriptsize had}}} <0,
P_{T,\mbox{\scriptsize antipar}} &=& \frac{1}{|\vec{P}_{T,\mbox{\scriptsize had}}|} \cdot \sum_{\vec{P}_{T,i} \cdot \vec{P}_{T,\mbox{\tiny had}} < 0} \vec{P}_{T,i} \cdot \vec{P}_{T,\mbox{\scriptsize had}}, \\
\end{eqnarray*}
where the sums are over  calorimeter deposits with polar angle $\theta
> 0.1$,    excluding    the    identified    positron.    The    ratio
$P_{T,\mbox{\scriptsize   antipar}}/P_{T,\mbox{\scriptsize  par}}$  is
used   in   the   final   selection   to   separate   one-jet   events
($P_{T,\mbox{\scriptsize  antipar}}/P_{T,\mbox{\scriptsize  par}} \sim
0$)      from      multi-jet      events      ($P_{T,\mbox{\scriptsize
antipar}}/P_{T,\mbox{\scriptsize par}} > 0$).

%%%%%%%%%%%%%%%%%%%%%%%%%%%%%%%%%%%%%%%%%%%%%%%%%%%%%%%%%%%%%%%%%%%%%%

\subsection{Preselection for $\nu$ final state}

Events with a  neutrino in the final state have  a topology similar to
CC DIS. The following selection cuts were applied in order to select a
sample of high-$Q^2$ CC events and suppress the non-$ep$ contribution:

\begin{itemize}
\item  $Z$-coordinate  of  the  event  vertex  compatible  with  an  $ep$
  interaction, $|Z_{\rm vtx}|<50$~cm;
\item no  reconstructed positron satisfying the same  criteria used in
  $e^{+}$ final-state preselection;
\item high missing transverse momentum, $P_{T,{\rm miss}}>20 \gev$,
  where $P_{T,{\rm miss}}$ is the missing transverse momentum as
  measured by the CAL;
\item      $0.2      <y_{\mbox{\scriptsize      JB}}<0.95$,      where
  $y_{\mbox{\scriptsize    JB}}$    is    reconstructed   using    the
  Jacquet-Blondel method~\cite{JB}.
\end{itemize}

The analogue  of Eq.~3  for the invariant  mass of  the $\nu$-hadronic
system  was derived assuming  that the  missing $P_{T}$  and $E-P_{Z}$
resulted from a single neutrino:

\begin{equation*}          
%M^{2}_{\nu X}=2E^{\mbox{\scriptsize beam}}_{e}\left(\sum_{i}(E+P_{Z})_{i}+P^{2}_{T, {\rm miss}}/(2E^{\mbox{\scriptsize beam}}_{e}\left(1-y_{\mbox{\scriptsize JB}})\right)\right).
M^{2}_{\nu X}=2E^{\mbox{\scriptsize beam}}_{e}\left(\sum_{i}(E+P_{Z})_{i}+ \frac{P^{2}_{T, {\rm miss}}}{2E^{\mbox{\scriptsize beam}}_{e} \cdot (1-y_{\mbox{\scriptsize JB}})}\right).
\end{equation*} 

%The sum  runs over  all the calorimeter  energy deposits with  a polar
%angle $>0.1$, to exclude contribution from the proton remnant.

The mass resolution varied between  $10\%$ and $3\%$ in the mass range
$100 - 280 \gev$. On  average, the mass was slightly underestimated at
high  masses ($1.5\%$  at $280  \gev$), while  the  agreement improved
towards  low  masses  ($<  1\%$   above  $120  \gev$).  After  the  CC
preselection cuts  $265$ events survived,  in good agreement  with the
expectation of  the SM MC  of $277^{+18}_{-21}$. The SM  prediction is
dominated  by CC  DIS events,  with a  small contribution  coming from
photoproduction processes.

Figure~4   shows   the   distributions   of   $P_{T,   {\rm   miss}}$,
$y_{\mbox{\scriptsize JB}}$, $\log_{10}(Q^2_{\mbox{\scriptsize JB}})$,
where  the  $Q^{2}_{JB}$ is  reconstructed  using the  Jacquet-Blondel
method,  and  $P_{T,\rm  antipar}/P_{T,\rm  par}$  for  data  and  MC;
reasonable agreement is observed for all the variables.

%%%%%%%%%%%%%%%%%%%%%%%%%%%%%%%%%%%%%%%%%%%%%%%%%%%%%%%%%%%%%%%%%%%%%%

\subsection{Final selection for $e^+$ final state}
\label{sec-final-e}

The final selection  for the channels with a  final-state positron was
designed to reduce further the  contamination from NC DIS by requiring
high-$Q^{2}$ and high-$y$ events. The following cuts were applied:

\begin{itemize}
\item $Q^{2}_{\mbox{\scriptsize {DA}}} > 3000 \gev^{2}$;
\item   
 $y_{\mbox{\scriptsize  DA}}   >  y_{\mbox{\scriptsize  cut}}$,  where
$y_{\mbox{\scriptsize  cut}}$  was  optimized  as a  function  of  the
reconstructed mass using the  SM MC and ranges between $0.7$ and
$0.4$ for  masses between $100$ and  $280 \gev$. This  cut exploits the
different $y$-dependence  of NC  DIS, steeply decreasing  as $y^{-2}$,
and of a  scalar resonance, which has a  substantial contribution from
large $y$.

\end{itemize}

Finally, a cut on $P_{T,\mbox{\scriptsize
antipar}}/P_{T,\mbox{\scriptsize par}}$ was used to produce two
samples enriched with either one-jet or multi-jet events. The $e$-J
($e$-MJ) final sample was defined requiring:

\begin{itemize}
\item $P_{t,\mbox{\scriptsize antipar}}/P_{t,\mbox{\scriptsize par}}<(>)~0.05$.
\end{itemize}

Signal  efficiencies were  evaluated by  generating samples  of signal
events using {\sc Susygen} for  different values of the MSSM or mSUGRA
parameters.  For the  $e$-J channel,  the efficiencies  ranged between
$10$ and  $45\%$ in the mass  range $100-260 \gev$,  decreasing to $20
\%$ at  $280 \gev$.  For the MSSM  scenario, the efficiencies  for the
$e$-MJ channel were in the range $5-25 \%$ for stop masses between 200
and  $280  \gev$,   depending  mainly  on  the  masses   of  stop  and
$\chi^{+}_{1}$; the efficiency decreased towards lower and higher stop
masses.  For the  mSUGRA  scenario, the  efficiencies  for the  $e$-MJ
channel were  in the range $5-15  \%$ in most of  the parameter space.
Table~1 shows good agreement between the number of selected events and
SM   expectation.  Figures~5a   and  5b   show  reconstructed   mass
distributions of data and SM  MC for the $e$-J and $e$-MJ preselection
and final samples. The data distributions are well described by the SM
simulation.

%%%%%%%%%%%%%%%%%%%%%%%%%%%%%%%%%%%%%%%%%%%%%%%%%%%%%%%%%%%%%%%%%%%%%%

\subsection{Final selection for $\nu$ final state}
\label{sec-final-nu}

In  order   to  enhance  stop  sensitivity  and    reduce  further the
contribution of CC DIS, the final selection required:

\begin{itemize}
\item $y_{\mbox{\scriptsize {JB}}} > 0.6$;
\item $P_{t,\mbox{\scriptsize  antipar}}/P_{t,\mbox{\scriptsize par}} > 0.1$.
\end{itemize}

For the MSSM  scenario, the efficiencies were in  the range $15-35 \%$
for stop  masses between 180 and  $280 \gev$, depending  mainly on the
masses  of   the  stop  and   $\tilde{\chi}^{0}_{1}$;  the  efficiency
decreased  towards  lower  and  higher  stop masses.  For  the  mSUGRA
scenario, the efficiencies were in the  range $5-20 \%$ in most of the
considered parameter  space. Table~1 shows good  agreement between the
number of selected events and the SM expectation.

Figure~5c shows  reconstructed mass distributions  of data and  SM MC
for  the $\nu$-MJ preselection  and final  samples. The  MC simulation
also in this case describes the data reasonably well.

%%%%%%%%%%%%%%%%%%%%%%%%%%%%%%%%%%%%%%%%%%%%%%%%%%%%%%%%%%%%%%%%%%%%%%
% ----------------------------------------------------------------------------
%       Results
% ----------------------------------------------------------------------------
%%%%%%%%%%%%%%%%%%%%%%%%%%%%%%%%%%%%%%%%%%%%%%%%%%%%%%%%%%%%%%%%%%%%%%

\section{Results}
\label{sec-results}

Since no  evidence for  stop production was  found, limits at  95\% CL
were  set using  a  Bayesian approach.  The  limits were  set for  two
different SUSY scenarios: the  unconstrained MSSM model and the mSUGRA
model (see Section~2).

%%%%%%%%%%%%%%%%%%%%%%%%%%%%%%%%%%%%%%%%%%%%%%%%%%%%%%%%%%%%%%%%%%%%%%

\subsection{Systematic uncertainties} \label{sec-syst}

In the calculation of the  upper limit on $\lprim$, several sources of
systematic  uncertainties were  considered.  The following  systematic
uncertainties on the SM background expectation were evaluated:
\begin{itemize} 
\item the uncertainty from  the proton PDFs,
  evaluated    using   the   procedure    suggested   by    the   CTEQ
  group~\cite{cteqsys}, was $\pm 4\%$ for $\nu$-MJ and $\pm 2\%$ for $e$-MJ 
and $e$-J;
\item the uncertainty  on the calorimeter energy
  scale of  $\pm 1\%$  ($\pm  2\%$) for  the electromagnetic  (hadronic) section 
  led to an uncertainty on the SM event rate of $\pm 5\%$ for $\nu$-MJ and $^{+1\%}_{-3\%}$ for $e$-MJ
and $e$-J;
\item the use of MEPS instead of {\sc Ariadne} to simulate the QCD cascade
led to an uncertainty of $-3\%$ for $\nu$-MJ and $-6\%$ for 
$e$-MJ and $e$-J;
\item  the uncertainty  in  the  integrated luminosity  measurement was  
$\pm 2.25\%$.
\end{itemize}
In  addition,  the  following  uncertainties  related  to  the  signal
simulation were considered:
\begin{itemize} 
\item the uncertainties in  the signal efficiency due to interpolation
  between different SUSY scenarios was $\pm 15\%$;
\item the theoretical  uncertainty on the signal cross  section due to
  the uncertainty in the $d$-quark parton density~\cite{cteqsys} in the proton
  varied from $\pm 3\%$ to $\pm 80\%$ for masses between $100$ and $280 \gev$.
\end{itemize}
%

%%%%%%%%%%%%%%%%%%%%%%%%%%%%%%%%%%%%%%%%%%%%%%%%%%%%%%%%%%%%%%%%%%%%%%

\subsection{Limits for the MSSM model}

Assuming  the MSSM model,  the upper  limits on  $\lambda'_{131}$ were
evaluated as a function of the  stop mass. A scan of the mass spectrum
in  $1 \gev$  steps was  performed using  a sliding  window of  $\pm 2
\sigma_{M_{\tilde{t}}}$ for  $M_{\ell X}  < 250 \gev$  ($\ell =  e$ or
$\nu$),  where $\sigma_{M_{\tilde{t}}}$ is  the stop  mass resolution.
For masses larger than $250  \gev$, where the SM background is smaller
and  the expected  signal width  larger, the  condition $M_{\ell  X} >
M_{\tilde{t}}-2 \sigma_{M_{\tilde{t}}}$ was applied.

At  each  stop mass,  the  $95\%$  CL  limit on  $\lambda'_{131}$  was
evaluated using, for each channel, the data events, the SM predictions
and  the signal  expectation for  the corresponding  mass  window. The
signal  cross section  was calculated  in the  NWA  (Eq.~2), including
initial-state radiation  for the incoming  positron~\cite{isr} and the
next-to-leading-order    QCD~\cite{nlo}    corrections,   using    the
CTEQ6\cite{cteqsys}  set  of  parton  densities, while  the  branching
ratios for the  different channels and MSSM scenarios  were taken from
the {\sc  Susygen} simulation. The  total likelihood was  evaluated as
the  product  of  the  Poissonian  likelihoods of  each  channel.  The
systematic uncertainties described in Section~6.1 were included in the
likelihood  function   assuming  Gaussian  probability   densities.  A
Bayesian approach assuming  a flat prior for the  signal cross section
was then used to produce the limits.

Figure~6 shows the  $95\%$ CL limit on $\lambda'_{131}$  as a function
of the stop mass for the range $-300 < \mu < 300 \gev$, $100 < M_{2} <
300 \gev$ and  $2< \tan{\beta} <50$. The limits for  masses up to $250
\gev$   improve   on    the   low-energy   constraints   from   atomic
parity-violation  (APV)~\cite{apv} measurements  (dashed line)  and do
not  depend   strongly  on  the  different  SUSY   scenarios.  The  H1
collaboration   obtained   similar  constraints~\cite{H1:SUSY}   using
similar SUSY scenarios.

%%%%%%%%%%%%%%%%%%%%%%%%%%%%%%%%%%%%%%%%%%%%%%%%%%%%%%%%%%%%%%%%%%%%%%

\subsection{Limits for the mSUGRA model}

For fixed values of $\lprim$, constraints on the mSUGRA parameters can
be set in the  plane ($m_{0}$,$m_{1/2}$), when tan$\beta$, $A_{0}$ and
the  sign  of $\mu$  are  fixed.  The  parameter $A_{0}$  enters  only
marginally at  the electroweak  scale and was  set to zero.  Limits at
95\% CL were evaluated using a scan of the reconstructed mass spectrum
and the same Bayesian approach as in the MSSM case.

Figure~7   shows   the   95\%   CL   excluded  area   in   the   plane
($m_{0}$,$m_{1/2}$)  for $\lprim$=0.3,  $\tan{\beta}=6$  and $\mu  <0$
(hatched area).  The dark region  corresponds to values  of parameters
where  no   REWSB  solution  is  possible,  while   the  light  region
corresponds to neutralino masses less than $30 \gev$, already excluded
by  LEP~\cite{epj:c19:397}.  The   dashed  lines  indicate  curves  of
constant  stop mass close  to the  border of  the excluded  area. Stop
masses can be excluded up to  $250 \gev$ for $m_{0}$ smaller than $240
\gev$. 
% New text about radiactive correction
The effects of the SUSY radiative corrections
on the sparticle mass spectrum is included in SUSPECT and have been
taken into account. Such effects increase the stop mass and consequently
worsen the limits especially at large $m_0$. For example the point
$m_0 = 200 \gev$, $m_{1/2} = 110 \gev$ is at the boundary of the
ZEUS exclusion region and corresponds to $M_{\tilde{t}} = 256 \gev$. The
same point, if SUSY radiative corrections are neglected, corresponds to
$M_{\tilde{t}} = 243 \gev$ and would be well inside the ZEUS excluded region.

A   scan   towards   large   $\tan{\beta}$  was   performed   assuming
$M=m_{0}=m_{1/2}$. Figure~8 shows  the limits on $M$ as  a function of
$\tan{\beta}$  for $\lprim$=0.3  and $\mu  <0$.  
%
% new version
The limit on $M$ slightly increases from 130 to $140 \gev$ in the range
 $6 < \tan{\beta} < 40$.
%
% old version
%The limit  on $M$  is
%almost constant  around $135 -  140 \gev$ up to  $\tan{\beta}\sim 35$.
%
For  larger  values,  it  drops   because  the  large  mixing  in  the
$\tilde{\tau}$ sector  results in a light  $\tilde{\tau_1}$ state into
which  the $\tilde{t}$ can  decay. The  efficiency for  detecting such
decay   is    low.
%%%%% Start correction added by Lorenzo
The effect of  the SUSY radiative corrections is  to slightly decrease
the overall  limit and to shift towards larger $\tan{\beta}$ the point
where  the  stau  branching   ratio  opens  up;  neglecting  radiative
corrections the limit drops at $\tan{\beta} \simeq 37$.
%%%%% End correction added by Lorenzo
The  H1 collaboration  obtained  comparable constraints~\cite{H1:SUSY}
using the same mSUGRA scenarios.

%%%%%%%%%%%%%%%%%%%%%%%%%%%%%%%%%%%%%%%%%%%%%%%%%%%%%%%%%%%%%%%%%%%%%%

\subsection{Comparison to results from other colliders}

Studies on stop  in $\rpv$ SUSY scenarios have  been performed both at
LEP~\cite{opal:stop,      aleph:stop,       l3:stop}      and      the
Tevatron~\cite{cdf:stop},  looking for the  production of  stop pairs.
LEP mass  limits for  the stop, in  the case  of $\lambda' >  0$, were
obtained  by  the  OPAL~\cite{opal:stop}  and  ALEPH~\cite{aleph:stop}
collaborations   and  are  in   the  range   $85-98  \gev$.   The  CDF
collaboration~\cite{cdf:stop}  set a  stop  mass limit  at $122  \gev$
assuming $\lambda'_{33k} > 0$ and a branching ratio $B(\tilde{t} \to b
\tau) = 1$.

A more interesting comparison  between HERA and Tevatron sensitivities
can  be  done by  looking  at  Tevatron  results for  leptoquark  (LQ)
production. The D0 collaboration published limits on leptoquark masses
as  a function  of the  branching ratio  $B(LQ  \to eq)$~\cite{D0:lq}.
Since  leptoquarks and squarks  have analogous  production mechanisms,
such  limits can  be  converted into  limits  on the  stop  mass as  a
function  of $\lambda'_{131}$\cite{cgm,lorenzo} and  directly compared
with the results of this analysis.  In the case of the MSSM scenarios,
D0  limits are competitive  with those  of HERA  only for  the largest
values  of $M_{2}$  and  $|\mu|$, where  the  $\rpv$ decay  $\tilde{t}
\rightarrow eq$  dominates due to  the large chargino mass.  For lower
values of $M_{2}$  or $|\mu|$, the gauge stop  decays are relevant and
the ZEUS limits improve over those from D0 for masses larger than $150
\gev$. Figure~9  shows the comparison  between ZEUS and D0  limits for
three different regions of  the unconstrained MSSM parameter space. In
the mSUGRA scenarios considered here, the gauge stop decays are always
relevant and thus  the ZEUS limits are more  stringent than those from
D0.

%%%%%%%%%%%%%%%%%%%%%%%%%%%%%%%%%%%%%%%%%%%%%%%%%%%%%%%%%%%%%%%%%%%%%%
% ----------------------------------------------------------------------------
%       Conclusions
% ----------------------------------------------------------------------------
%%%%%%%%%%%%%%%%%%%%%%%%%%%%%%%%%%%%%%%%%%%%%%%%%%%%%%%%%%%%%%%%%%%%%%

\section{Conclusions}

A  search  for  stop  production  in $e^+p$  collisions  at  HERA  was
performed  using  an   integrated  luminosity  of  $65$~pb$^{-1}$.  No
evidence was  found for resonances  in the decay channels  with jet(s)
and  one high-$P_{T}$  positron  or neutrino.  The  results have  been
interpreted in the framework of the $R$-parity-violating MSSM, setting
limits on  the Yukawa coupling  $\lambda'_{131}$ as a function  of the
stop  mass.  These  limits  exhibit  a weak  dependence  on  the  MSSM
parameters $\mu$, $M_{2}$ and $\tan{\beta}$ and improve on limits from
Tevatron in  a large  part of the  considered parameter space,  and on
limits from  low-energy atomic parity-violation  measurements for stop
masses lower than $250 \gev$. Direct limits on the stop mass have also
been derived  within the  mSUGRA model. In  this model only  five free
parameters determine  the full  supersymmetric mass spectrum.  In this
case,  for $\lprim =0.3$,  $\tan{\beta}=6$, $\mu  < 0$  and $A_{0}=0$,
stop with masses  as high as $260 \gev$ are excluded  for a large part
of the parameter space.

%%%%%%%%%%%%%%%%%%%%%%%%%%%%%%%%%%%%%%%%%%%%%%%%%%%%%%%%%%%%%%%%%%%%%%
% ----------------------------------------------------------------------------
%       Acknowledgements
% ----------------------------------------------------------------------------
%%%%%%%%%%%%%%%%%%%%%%%%%%%%%%%%%%%%%%%%%%%%%%%%%%%%%%%%%%%%%%%%%%%%%%

\section*{Acknowledgements} 

 It  is a  pleasure to  thank the  DESY Directorate  for  their strong
support  and encouragement.  The remarkable  achievements of  the HERA
machine  group were essential  for the  successful completion  of this
work  and  are  greatly  appreciated.  The  design,  construction  and
installation  of the  ZEUS  detector  has been  made  possible by  the
efforts of many people who are not listed as authors.
%%%%%%%%%%%%%%%%%%%%%%%%%%%%%%%%%%%%%%%%%%%%%%%%%%%%%%%%%%%%%%%%%%%%%%%%%%%%%%%
%------------------------------------------------------------------------------
%       Bibliography
%------------------------------------------------------------------------------
%\include{DESY-06-144-ref}
%%%%%%%%%%%%%%%%%%%%%%%%%%%%%%%%%%%%%%%%%%%%%%%%%%%%%%%%%%%%%%%%%%%%%%%%%%%%%%%
\providecommand{\etal}{et al.\xspace}
\providecommand{\coll}{Coll.\xspace}
\catcode`\@=11
\def\@bibitem#1{%
\ifmc@bstsupport
  \mc@iftail{#1}%
    {;\newline\ignorespaces}%
    {\ifmc@first\else.\fi\orig@bibitem{#1}}
  \mc@firstfalse
\else
  \mc@iftail{#1}%
    {\ignorespaces}%
    {\orig@bibitem{#1}}%
\fi}%
\catcode`\@=12

%%%%%%%%%%%%%%%%%%%%%%%%%%%%%%%%%%%%%%%%%%%%%%%%%%%%%%%%%%%%%%%%%%%%%%%%%%%%%%%
%------------------------------------------------------------------------------
%       Tables
%------------------------------------------------------------------------------
%\include{DESY-06-144-tab}
%%%%%%%%%%%%%%%%%%%%%%%%%%%%%%%%%%%%%%%%%%%%%%%%%%%%%%%%%%%%%%%%%%%%%%%%%%%%%%%
%-------------------------------------------------------------------------------
%       An example table
%-------------------------------------------------------------------------------
%
\newpage
\begin{table}
\label{tab-sum}
\begin{center}%
\begin{tabular}{|c|c|c|c|c|c|c|} 
 \hline
 Channel & $Q^{2}_{\mbox{\scriptsize DA}}$ (GeV$^{2}$) & $y_{\mbox{\scriptsize cut}}(M_{\tilde{t}})$ & $P_{T,\mbox{\scriptsize antipar}}/P_{T,\mbox{\scriptsize par}}$ & Data & SM & Eff. MSSM \\ \hline \hline 
%$e$-J      & $0.4-0.7$ & $<0.05$ & 85 & $74.5^{+2.6}_{-7.2}$  & 0.3       & 0.3   \\ \hline
%$e$-MJ     & $0.4-0.7$ & $>0.05$ & 63 & $58.8^{+2.1}_{-9.0}$  & 0.15-0.2  & 0.14  \\ \hline 
%$\nu$-MJ   & $0.6$     & $>0.1$  & 19 & $20.9^{+1.7}_{-2.9}$  & 0.15-0.35 & 0.19  \\ \hline %\hline
$e$-J      & $>3000$ &  $0.4-0.7$ & $<0.05$ & 85 & $74.5^{+3.5}_{-6.0}$  & 0.3        \\ \hline
$e$-MJ     & $>3000$ &  $0.4-0.7$ & $>0.05$ & 63 & $58.8^{+3.0}_{-5.0}$  & 0.15-0.2   \\ \hline 
$\nu$-MJ   & $-$     &  $0.6$     & $>0.1$  & 19 & $20.9^{+1.5}_{-1.6}$  & 0.15-0.35  \\ \hline %\hline
\end{tabular}
\end{center}
\caption{Summary  of  final selection  cuts,  number  of observed  and
expected  events for the SM and signal  efficiencies for  MSSM ($M_{\sTop}=220$~GeV)
 for the different channels discussed in the text.}

\end{table}
%
%%%%%%%%%%%%%%%%%%%%%%%%%%%%%%%%%%%%%%%%%%%%%%%%%%%%%%%%%%%%%%%%%%%%%%%%%%%%%%%
%------------------------------------------------------------------------------
%       Figures
%------------------------------------------------------------------------------
%\include{DESY-06-144-fig}
%%%%%%%%%%%%%%%%%%%%%%%%%%%%%%%%%%%%%%%%%%%%%%%%%%%%%%%%%%%%%%%%%%%%%%%%%%%%%%%
%-------------------------------------------------------------------------------
%       Results
%-------------------------------------------------------------------------------
%
%%%%%%%%%%%%%%%%%%%%%%%%%%%%%%%%%%%%%%%%%%%%%%%%%%%%%%%%%%%%%%%%%%%%%%%%%%%%%%%%
% Feynman diagrams
%
\begin{figure}[p]
\centering
\mbox{\epsfig{file=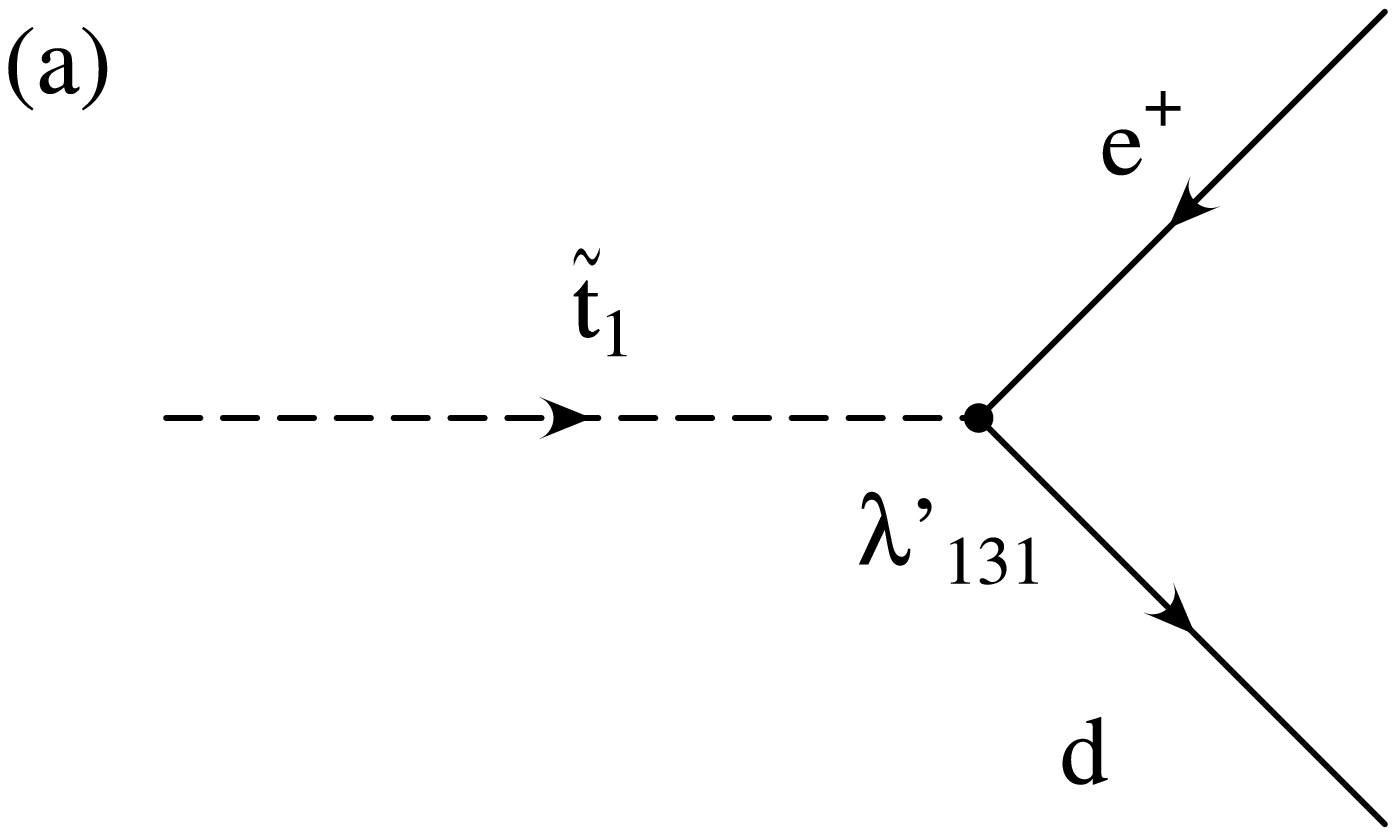,width=7.7cm}} \\
\mbox{\epsfig{file=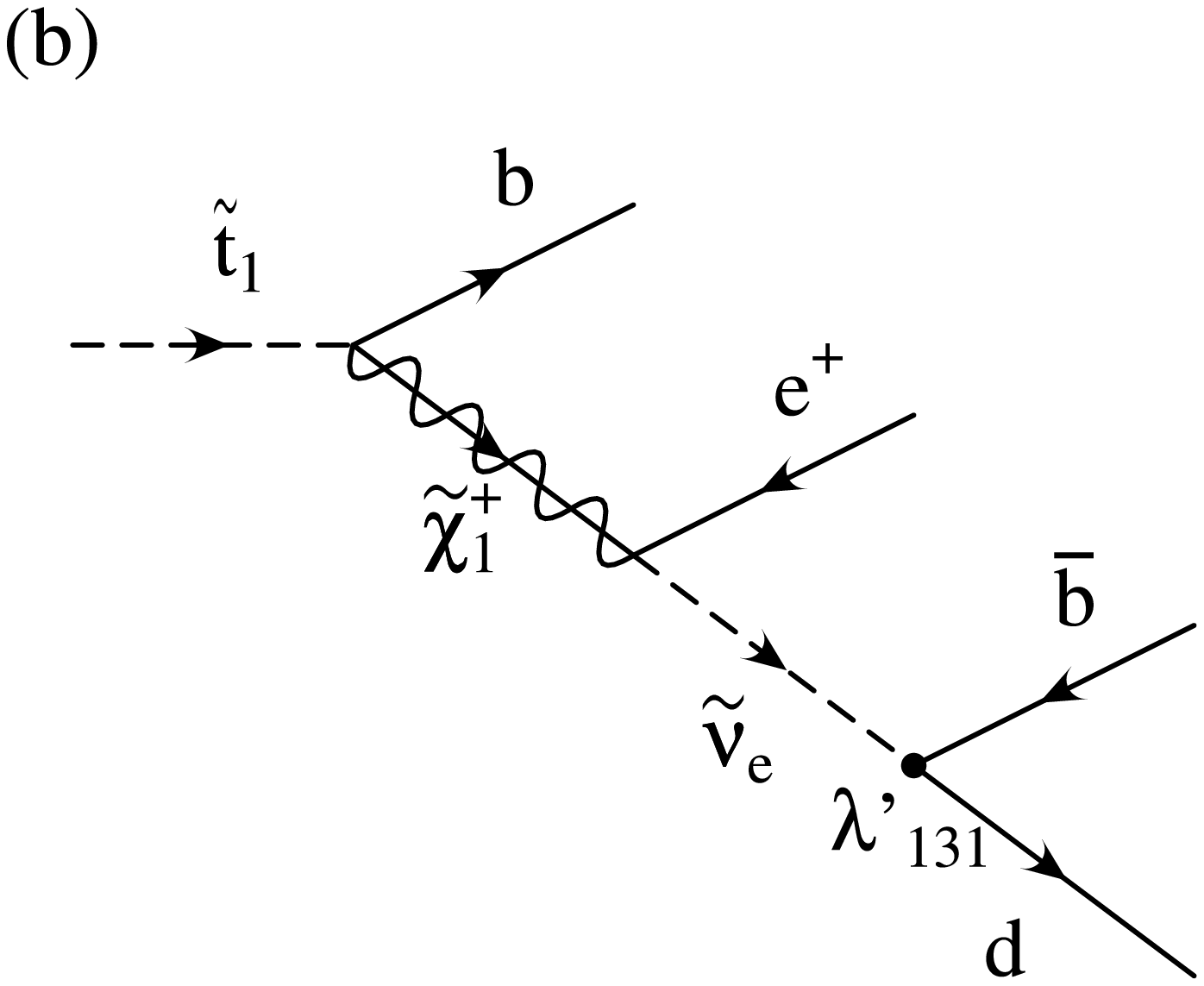,width=7.5cm}} \\
\mbox{\epsfig{file=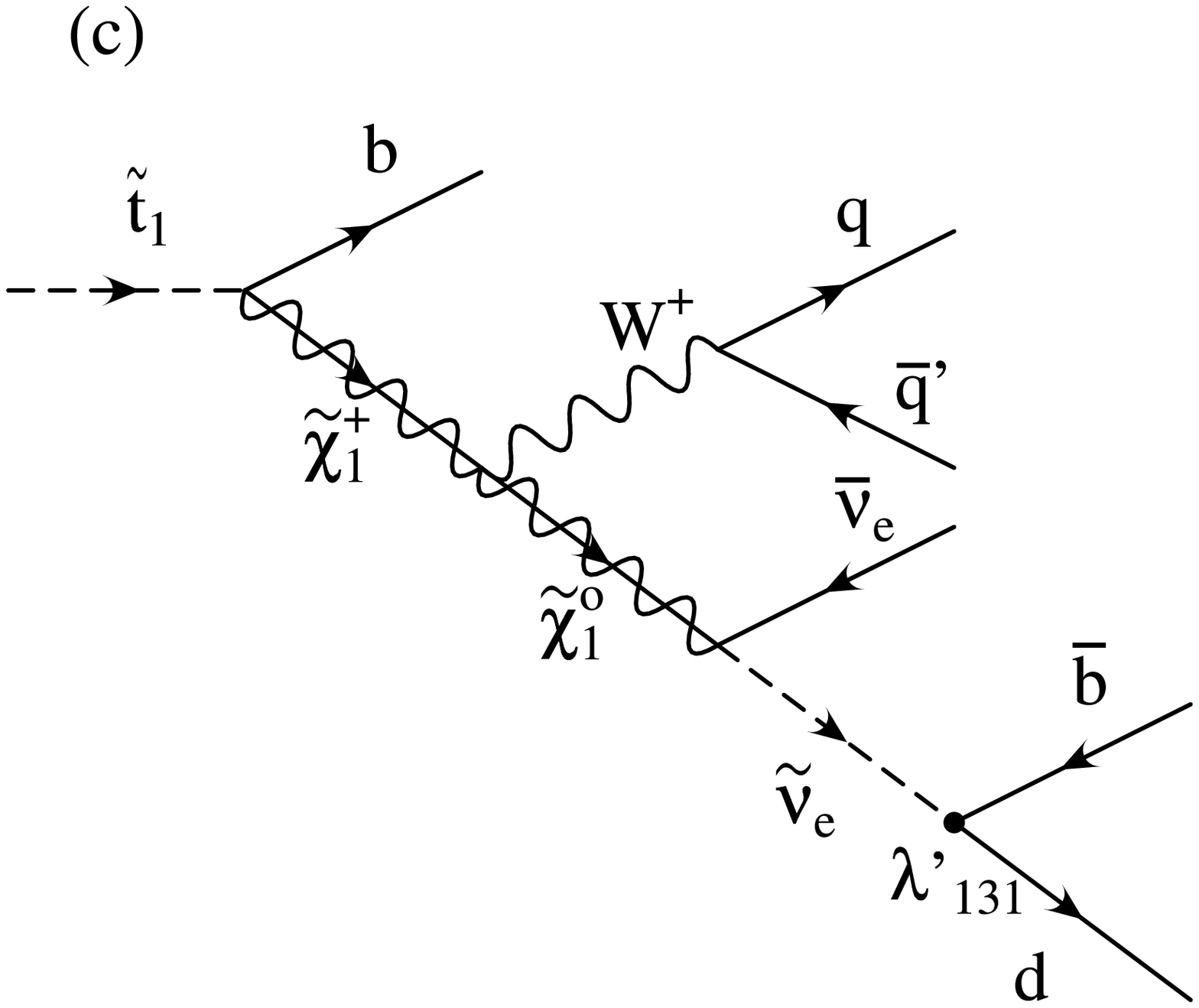,width=9.5cm}} 
\caption{Considered decay modes of the stop squark: the e-J channel (a), the e-MJ channel (b) and the $\nu$-MJ channel (c).}
\label{fig-diagrams}
\end{figure}
%
%%%%%%%%%%%%%%%%%%%%%%%%%%%%%%%%%%%%%%%%%%%%%%%%%%%%%%%%%%%%%%%%%%%%%%%%%%%%%%%%
% 
%
\begin{figure}[p]
\vfill
\begin{center}
\mbox{\epsfig{file=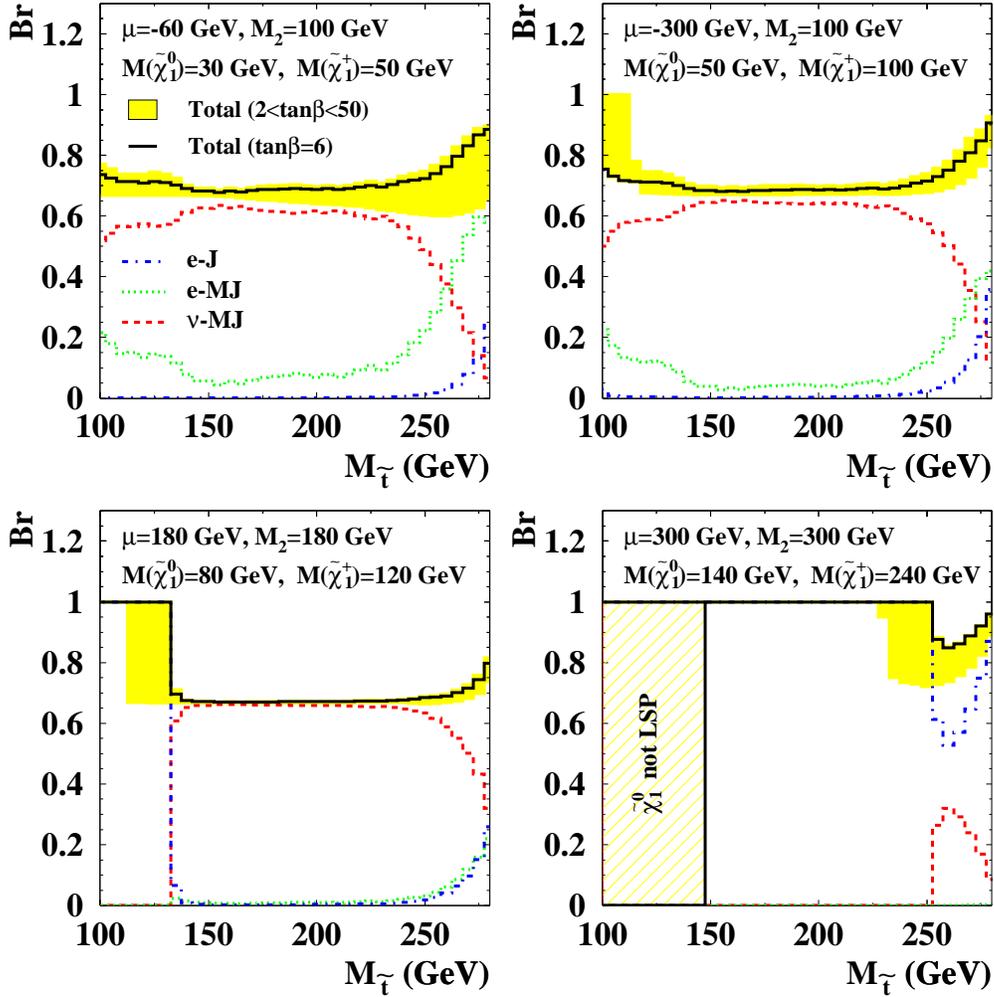,width=15cm}} \hfill
\end{center}
\caption{
Branching ratios  as a function of  the stop mass  for different $\mu$
and $M_{2}$ values,  and for $\lambda'_{131}$ equal to  the ZEUS limit
for  the considered  scenarios. 
The three channels considered in the analysis and their sum are shown for $\tan{\beta}=6$.
The shaded band is  the range of
the total branching ratio for $2<\tan{\beta}<50$. 
}

\label{fig-2}
\vfill
\end{figure}
%
%%%%%%%%%%%%%%%%%%%%%%%%%%%%%%%%%%%%%%%%%%%%%%%%%%%%%%%%%%%%%%%%%%%%%%%%%%%%%%%%
% Control plots NC selection
%
\begin{figure}[p]
\vfill
\begin{center}
\mbox{\epsfig{file=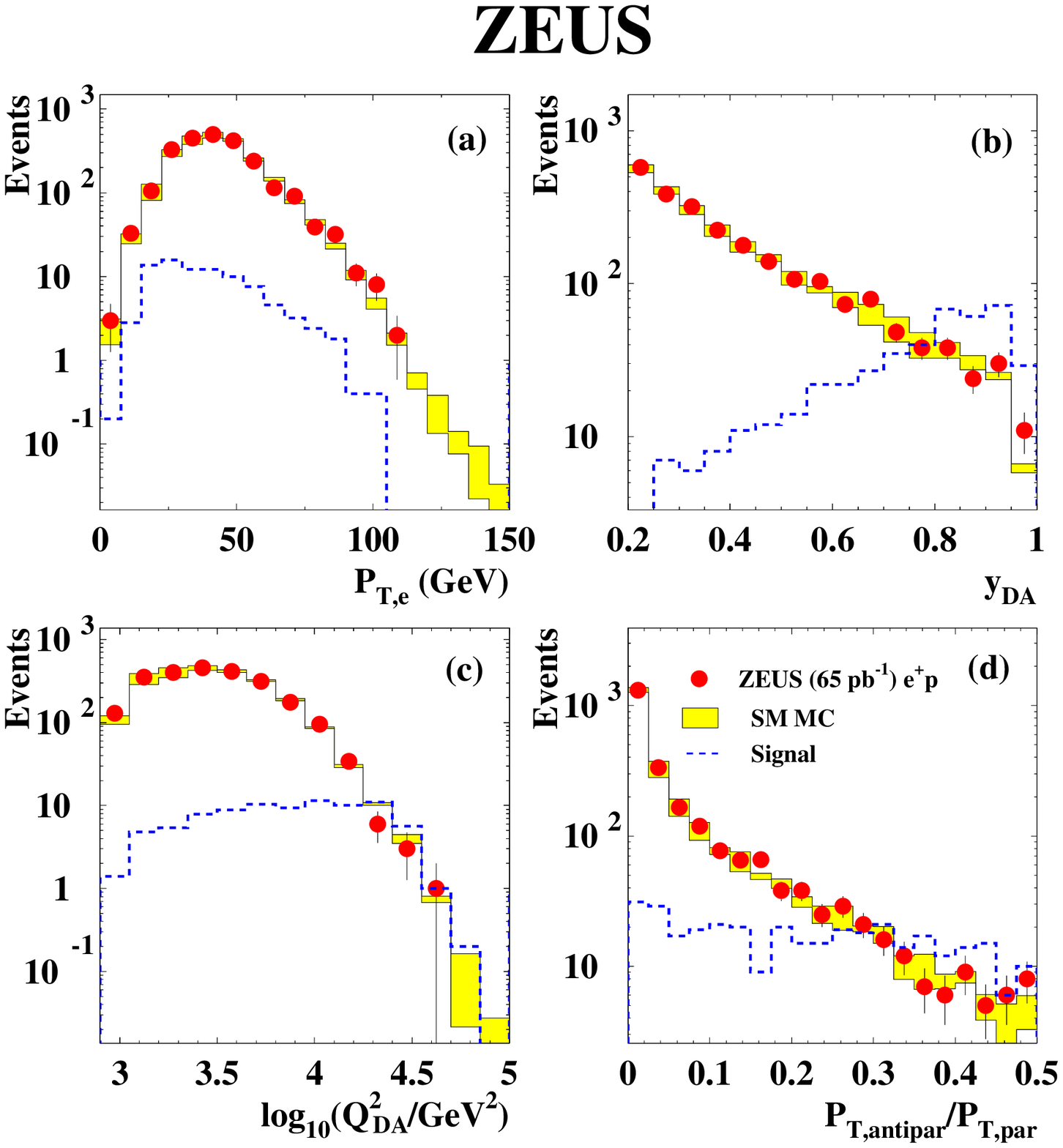,width=15cm}} \hfill
\end{center}
\caption{
Comparison between  data (dots) and SM MC  (histograms) for (a) $P_{T,e}$,
(b) $y_{\mbox{\scriptsize    {DA}}}$,  (c)  $\log_{10}(Q^2_{\mbox{\scriptsize
DA}})$  and (d) $P_{T,\mbox{\scriptsize  antipar}}/P_{T,\mbox{\scriptsize
par}}$ after the preselection for $e^{+}$ final state. 
The  band  represents   the  SM expectation with its uncertainty.  
The MSSM  signal  for the e-MJ channel (dashed  line)  with $M_{\tilde{t}}=220
\gev$,  $M_{2}=100  \gev$  and  $\mu=-300  \gev$ is  also  shown,  with
arbitrary normalization.}
\label{fig-ccnc}
\vfill
\end{figure}
%
%%%%%%%%%%%%%%%%%%%%%%%%%%%%%%%%%%%%%%%%%%%%%%%%%%%%%%%%%%%%%%%%%%%%%%%%%%%%%%%%
% Control plots CC selection
%
\begin{figure}[p]
\vfill
\begin{center}
\mbox{\epsfig{file=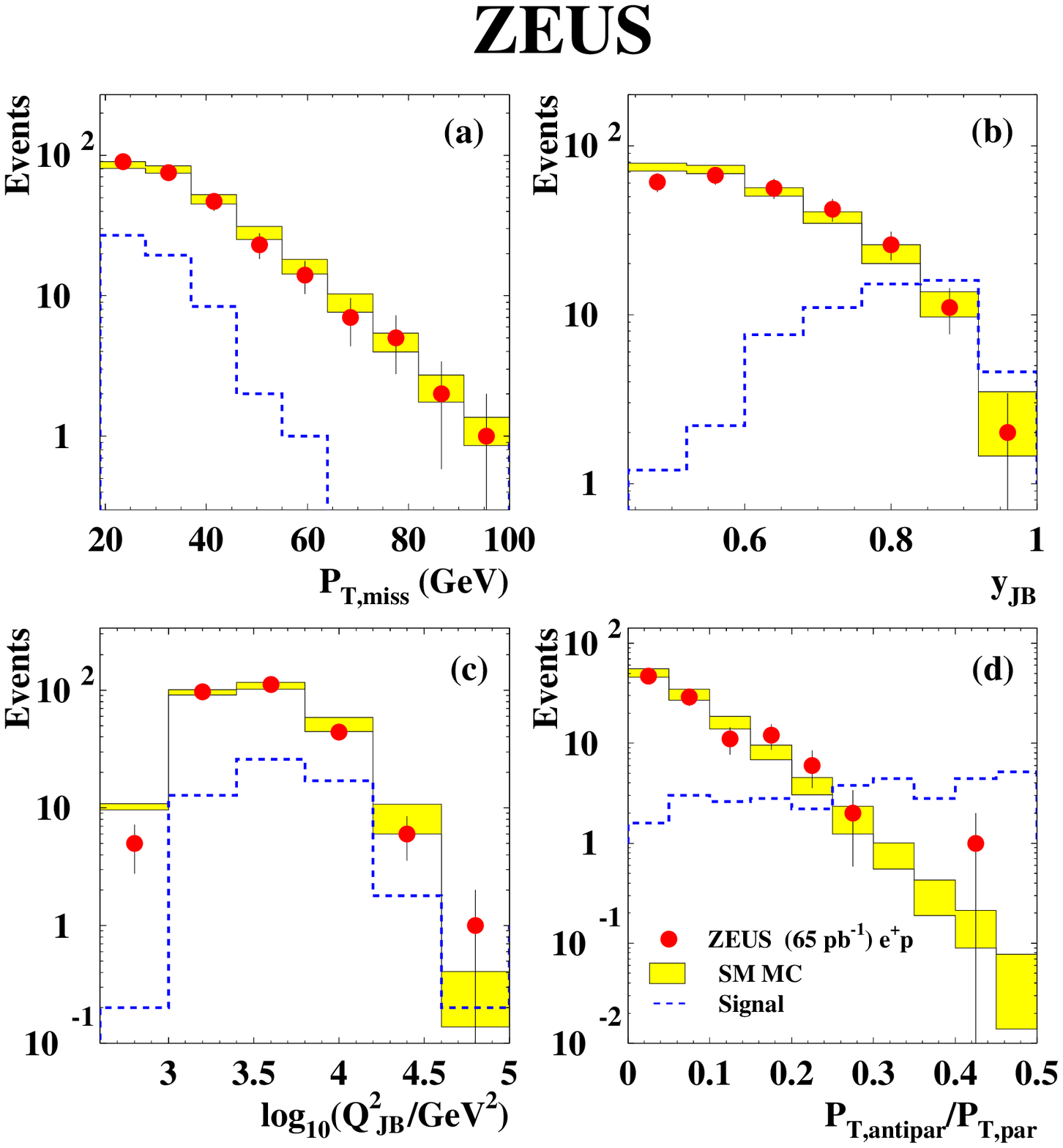,width=15cm}} \hfill
\end{center}
\caption{
Comparison  between data (dots)  and SM  MC (histograms)  for (a) $P_{T, \mbox{\scriptsize miss}}$,
(b) $y_{\mbox{\scriptsize  JB}}$, (c) $\log_{10}(Q^2_{\mbox{\scriptsize JB}})$
and  (d) $P_{t,\mbox{\scriptsize  antipar}}/P_{t,\mbox{\scriptsize  par}}$ after the preselection for $\nu$ final state.
The  band  represents   the  SM expectation with its uncertainty.  
The MSSM  signal (dashed   line)  with  $M_{\tilde{t}}=220  \gev$, 
$M_{2}=100  \gev$ and  $\mu=-300 \gev$  is also  shown,  with arbitrary
normalization.}
\label{fig-cccc}
\vfill
\end{figure}
%
%%%%%%%%%%%%%%%%%%%%%%%%%%%%%%%%%%%%%%%%%%%%%%%%%%%%%%%%%%%%%%%%%%%%%%%%%%%%%%%%
% Mass distribution
%
\begin{figure}[p]
\vfill
\begin{center}
\mbox{\epsfig{file=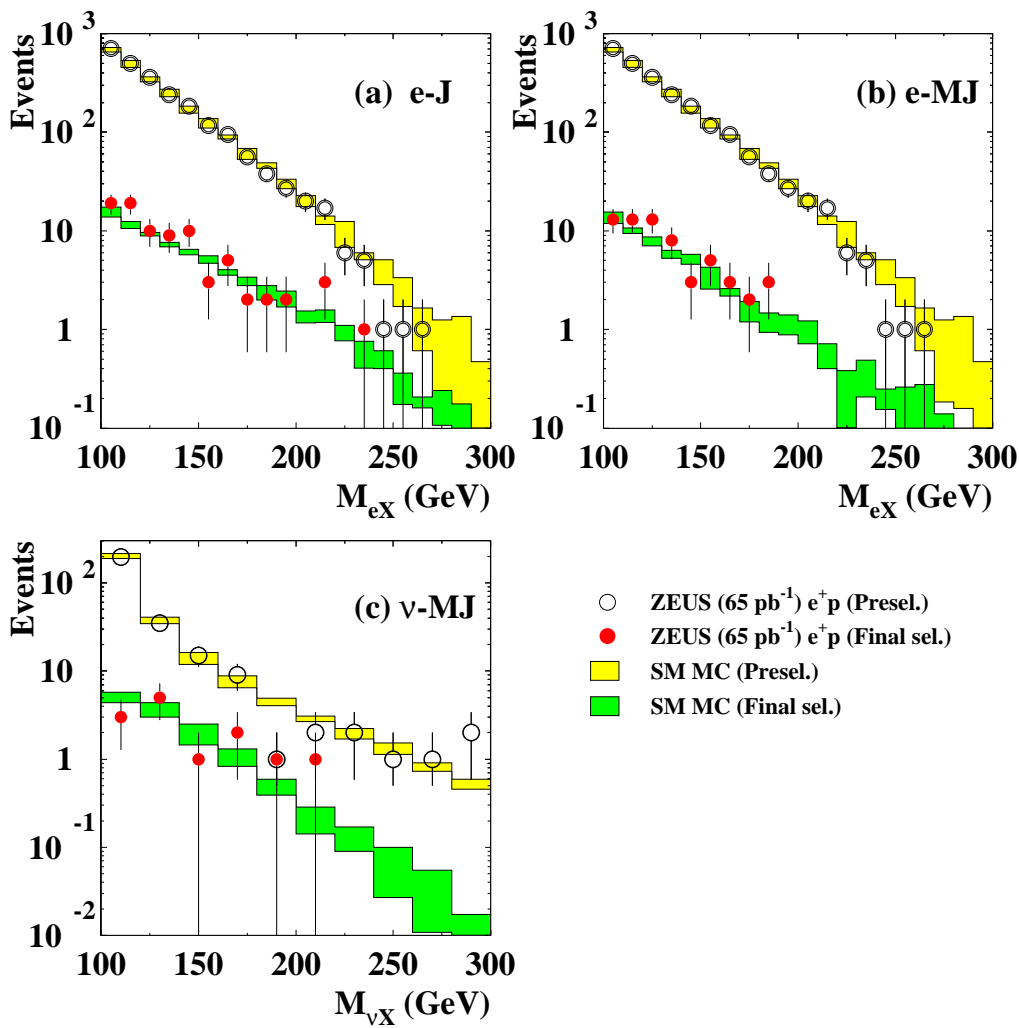,width=15cm}} \hfill
\end{center}
\caption{
Reconstructed mass for (a)  e-J, (b) e-MJ and  (c) $\nu$-MJ channels.
The data (dots)  and SM MC (histograms) after  preselection (empty circles
and  light  histograms)  and  final  selection (filled  circles  and  dark
histograms) are shown. 
The  band  represents   the  SM expectation with its uncertainty.}
\label{fig-mass}
\vfill
\end{figure}
%
%%%%%%%%%%%%%%%%%%%%%%%%%%%%%%%%%%%%%%%%%%%%%%%%%%%%%%%%%%%%%%%%%%%%%%%%%%%%%%%%
% MSSM Limits
%
\begin{figure}[p]
\vfill
\begin{center}
\mbox{\epsfig{file=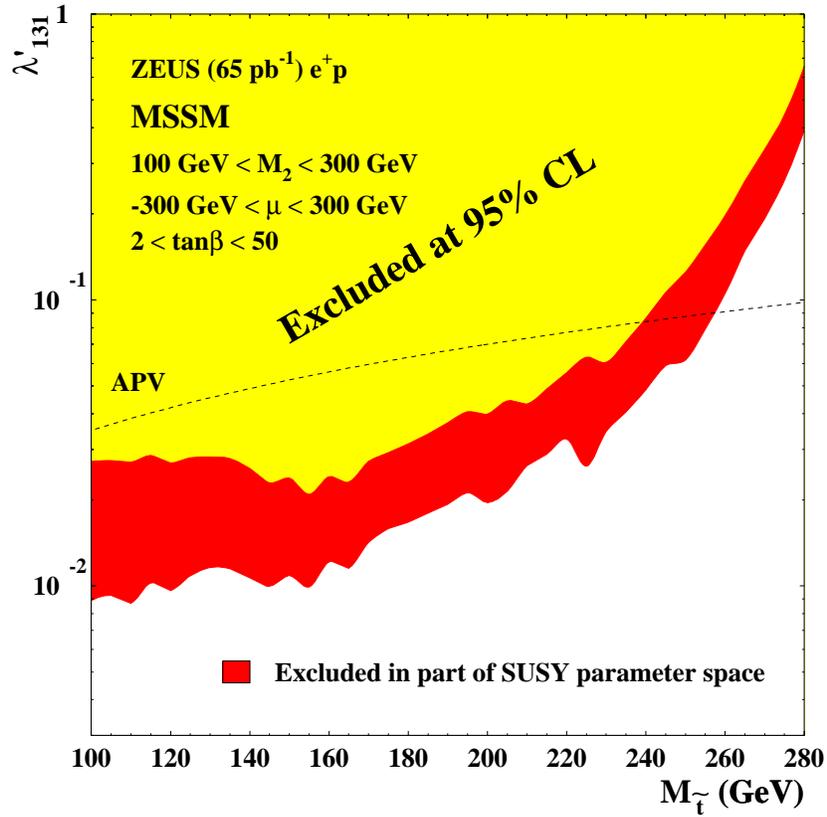,width=12cm}} \hfill
\end{center}
\caption{Exclusion limits on $\lambda'_{131}$ as a function of the stop mass for the MSSM model.
  The light (dark) region is excluded in all (part of) the considered
  SUSY parameter space. 
  The region above the dashed line is excluded by low-energy atomic parity-violation (APV) measurements.
}
\label{fig-mssm}
\vfill
\end{figure}
%
%%%%%%%%%%%%%%%%%%%%%%%%%%%%%%%%%%%%%%%%%%%%%%%%%%%%%%%%%%%%%%%%%%%%%%%%%%%%%%%%
% mSUGRA Limits
%
\begin{figure}[p]
\vfill
\begin{center}
\mbox{\epsfig{file=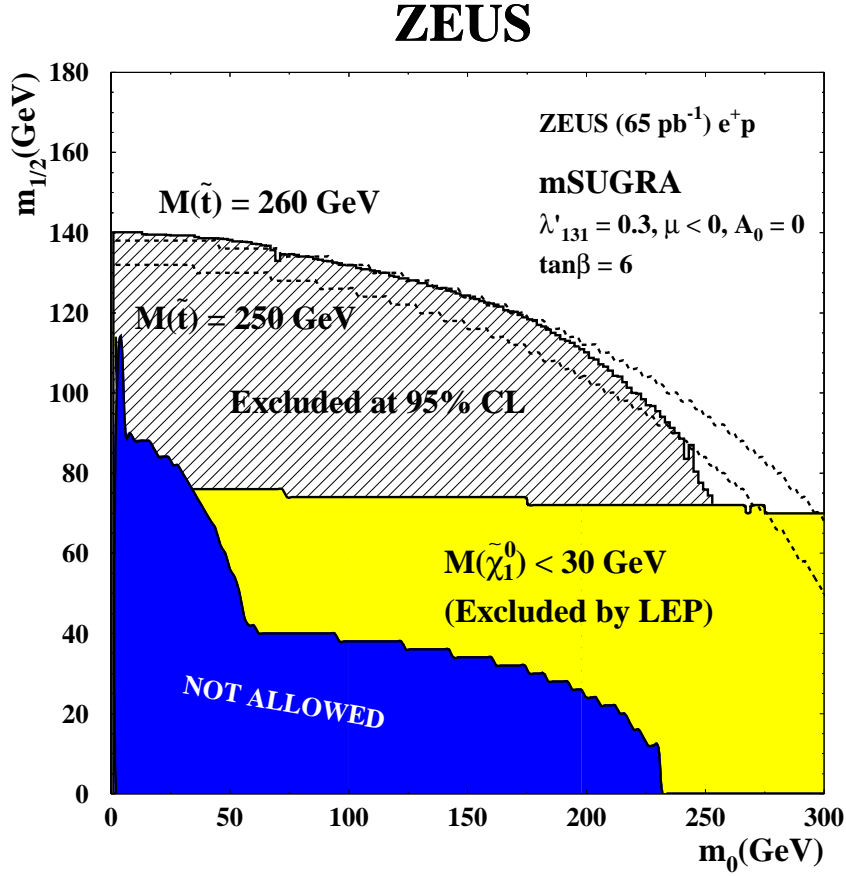,width=12cm}} \hfill
\end{center}
\caption{ Exclusion  limits for  mSUGRA with tan$\beta$=6,
$\lprim$=0.3, $\mu < 0$  and $A_{0}=0$ (hatched area). The dark-shaded region corresponds to
values of parameters where no radiative electroweak symmetry-breaking solution is
possible. The light-shaded region corresponds to neutralino masses (LSP)
less than  $30 \gev$, already excluded  by LEP results.  The dashed lines
indicate the curve of constant stop mass close to the border of the excluded
region.}
\label{fig-msugra}
\vfill
\end{figure}
%
%%%%%%%%%%%%%%%%%%%%%%%%%%%%%%%%%%%%%%%%%%%%%%%%%%%%%%%%%%%%%%%%%%%%%%%%%%%%%%%%
% mSUGRA limits on tan(b)
%
\begin{figure}[p]
\vfill
\begin{center}
\mbox{\epsfig{file=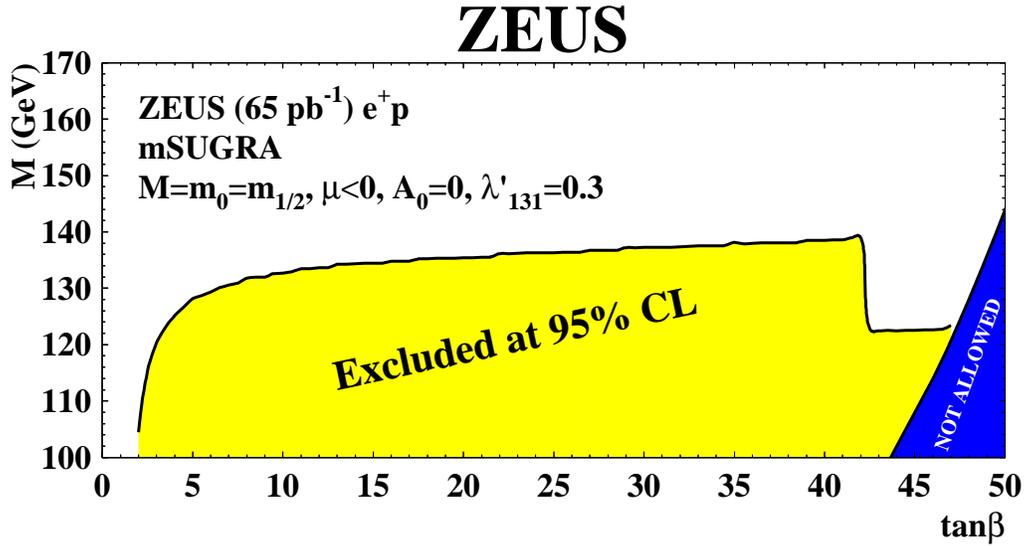,width=15cm}} \hfill
\end{center}
\caption{
Exclusion   limit    for   mSUGRA   on   the    mass   parameter   $M$
($M=m_{0}=m_{1/2}$)   as   a   function   of   $\tan{\beta}$   for
$\lprim$=0.3, $\mu <0$ and $A_{0}=0$. The dark-shaded region corresponds to
values of parameters where no radiative electroweak symmetry-breaking solution is
possible.}
\label{fig-8}
\vfill
\end{figure}
%
%%%%%%%%%%%%%%%%%%%%%%%%%%%%%%%%%%%%%%%%%%%%%%%%%%%%%%%%%%%%%%%%%%%%%%%%%%%%%%%%
% 
%
\begin{figure}[p]
\vfill
\begin{center}
\mbox{\epsfig{file=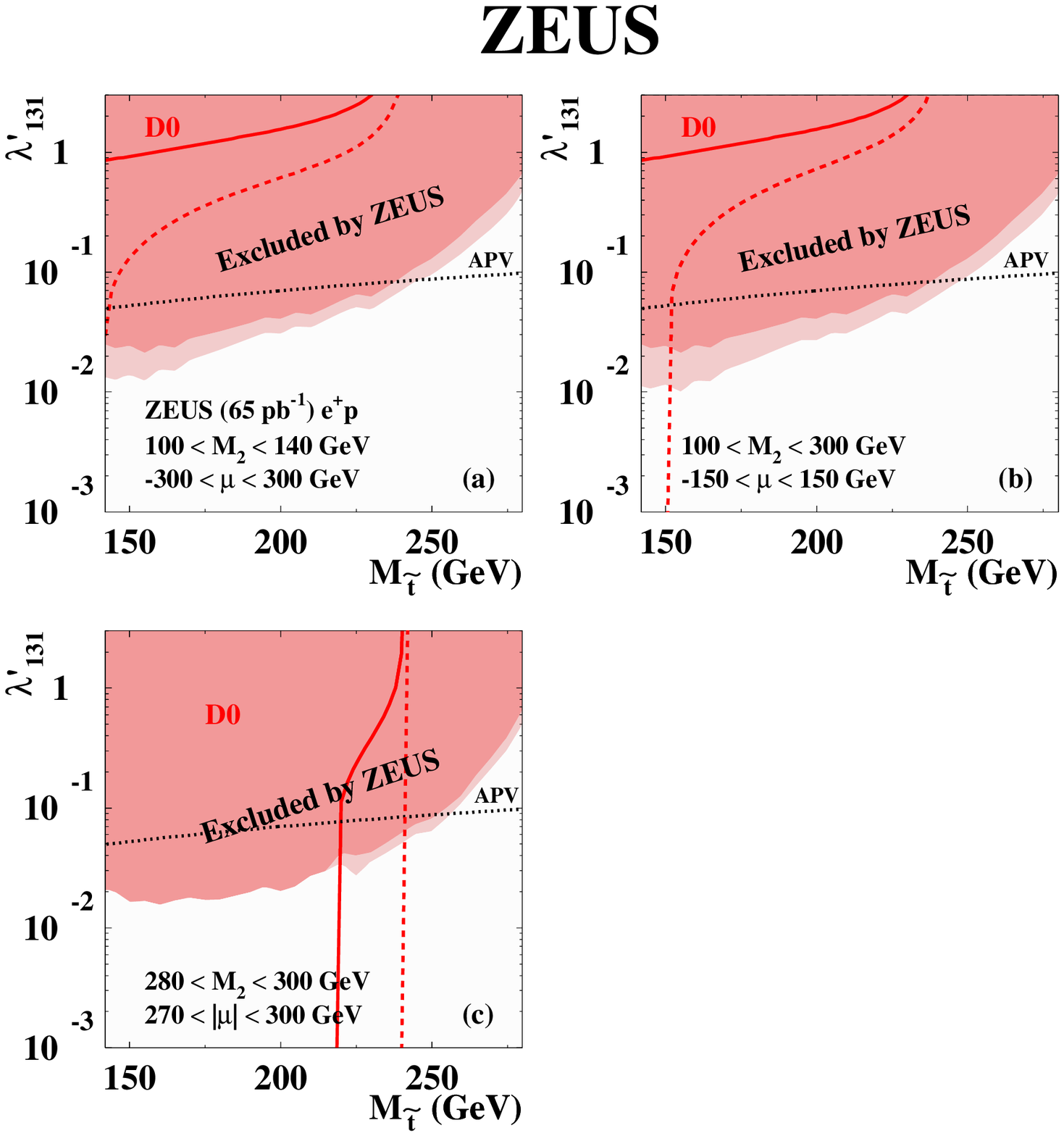,width=15cm}} \hfill
\end{center}
\caption{Comparison between  ZEUS, D0 and atomic parity violating (APV) limits in the MSSM scenario for $\tan{\beta}=6$ and (a) low $M_{2}$,  (b) low $|\mu|$   and  (c) high  $M_{2}$  and $|\mu|$.
The regions excluded by ZEUS and D0 are shown by the dark-shaded area and by the area above the full line, respectively.
The regions excluded in part of the parameter space are the light-shaded area (ZEUS) and the area between the full and the dashed line (D0). 
The region of exclusion for  atomic
parity-violation (APV) is above the dotted line.
}
\label{fig-9}
\vfill
\end{figure}
%
%%%%%%%%%%%%%%%%%%%%%%%%%%%%%%%%%%%%%%%%%%%%%%%%%%%%%%%%%%%%%%%%%%%%%%%%%%%%%%%%
%%%%%%%%%%%%%%%%%%%%%%%%%%%%%%%%%%%%%%%%%%%%%%%%%%%%%%%%%%%%%%%%%%%%%%%%%%%%%%%
%
%       ... that's it
%
\end{document}

%% file: zeus_def.tex
\newcommand{\ZcoosysB}{%
The ZEUS coordinate system is a right-handed Cartesian system, with the $Z$
axis pointing in the proton beam direction, referred to as the ``forward
direction'', and the $X$ axis pointing left towards the centre of HERA.
The coordinate origin is at the nominal interaction point.\xspace}
\newcommand{\Zpsrap}{%
\xspace}

\newcommand{\ZcoosysfnBeta}{\footnote{\ZcoosysB\Zpsrap}}
\newcommand{\Zdetdesc}{%
A detailed description of the ZEUS detector can be found 
elsewhere~\cite{zeus:1993:bluebook}. A brief outline of the 
components that are most relevant for this analysis is given
below.\xspace}
\newcommand{\Zctddesc}[1]{%
Charged particles are tracked in the central tracking detector (CTD)~\citeCTD,
which operates in a magnetic field of $1.43\Tesla$ provided by a thin 
superconducting coil. The CTD consists of 72~cylindrical drift chamber 
layers, organized in 9~superlayers covering the polar-angle#1 region 
\mbox{$15^\circ<\theta<164^\circ$}. The transverse-momentum resolution for
full-length tracks is $\sigma(p_T)/p_T=0.0058p_T\oplus0.0065\oplus0.0014/p_T$,
with $p_T$ in $\Gev$.}
\newcommand{\Zcaldesc}{%
The high-resolution uranium--scintillator calorimeter (CAL)~\citeCAL consists 
of three parts: the forward (FCAL), the barrel (BCAL) and the rear (RCAL)
calorimeters. Each part is subdivided transversely into towers and
longitudinally into one electromagnetic section (EMC) and either one (in RCAL)
or two (in BCAL and FCAL) hadronic sections (HAC). The smallest subdivision of
the calorimeter is called a cell.  The CAL energy resolutions, as measured under
test-beam conditions, are $\sigma(E)/E=0.18/\sqrt{E}$ for electrons and
$\sigma(E)/E=0.35/\sqrt{E}$ for hadrons, with $E$ in $\Gev$.}
\chardef\usc=95
\chardef\til=126
\catcode`\@=11 % @ signs are now treated as letters
\DeclareRobustCommand\xdotspace{\futurelet\@let@token\@xdotspace}
\def\@xdotspace{%
  \ifx\@let@token.\else
  \ifx\@let@token\bgroup.\else
  \ifx\@let@token\egroup.\else
  \ifx\@let@token\/.\else
  \ifx\@let@token\ .\else
  \ifx\@let@token~.\else
  \ifx\@let@token!.\else
  \ifx\@let@token,.\else
  \ifx\@let@token:.\else
  \ifx\@let@token;.\else
  \ifx\@let@token?.\else
  \ifx\@let@token/.\else
  \ifx\@let@token'.\else
  \ifx\@let@token).\else
  \ifx\@let@token-.\else
  \ifx\@let@token\@xobeysp.\else
  \ifx\@let@token\space.\else
  \ifx\@let@token\@sptoken.\else
   .\space
   \fi\fi\fi\fi\fi\fi\fi\fi\fi\fi\fi\fi\fi\fi\fi\fi\fi\fi}
\catcode`\@=12 % @ signs are no longer letters

\newcommand{\stru}[2]{%
   \relax\ifmmode\hbox{\vrule height#1 depth#2 width0pt}%
   \else\vrule height#1 depth#2 width0pt\fi}
\newcommand{\Ronum}[1]{\uppercase\expandafter{\romannumeral#1}}
\newcommand{\ronum}[1]{\expandafter{\romannumeral#1}}
\DeclareRobustCommand{\LaTeXZ}{%
  \LaTeX\kern-.05em4\kern-.1em
  {\raisebox{-0.2ex}{$\scriptstyle\text{ZEUS}$}}\xspace}
\DeclareMathAlphabet{\mathbf}{OT1}{cmr}{bx}{sl}
\newcommand{\eVdist}{\kern-0.06667em}

\newcommand{\Gev}{{\text{Ge}\eVdist\text{V\/}}}

\newcommand{\gev}{{\,\text{Ge}\eVdist\text{V\/}}}

\newcommand{\Tesla}{\,\text{T}}

\newcommand{\slashfrac}[2]{%
  \raisebox{0.5ex}{\ensuremath #1}\kern-0.12em/\kern-0.08em
  \raisebox{-.8ex}{\ensuremath #2}}
\newcommand{\sqr}[3]{%
    {\vcenter{\hrule height.#3ex\hbox{\vrule width.#2ex height#1ex
     \kern#1ex\vrule width.#3ex}\hrule height.#2ex}}}

\catcode`\@=11 % @ signs are now treated as letters
\newcommand{\parenbar}{\mathpalette\p@renb@r}
\def\p@renb@r#1#2{\vbox{%
  \ifx#1\scriptscriptstyle \dimen@.7em\dimen@ii.2em\else
  \ifx#1\scriptstyle \dimen@.8em\dimen@ii.25em\else
  \dimen@1em\dimen@ii.4em\fi\fi \offinterlineskip
  \ialign{\hfill##\hfill\cr
    \vbox{\hrule width\dimen@ii}\cr
    \noalign{\vskip-.3ex}%
    \hbox to\dimen@{$\mathchar300\hfil\mathchar301$}\cr
    \noalign{\vskip-.3ex}%
    $#1#2$\cr}}}
\catcode`\@=12 % @ signs are no longer letters

\newcommand{\IP}{{\rm I$\kern-0.01667em$P}\xspace}

\mathchardef\qsm=63
\mathchardef\pls=43
\mathchardef\mns=512
\mathchardef\plm=518
\mathchardef\eql=61
\mathchardef\smallleft=300
\mathchardef\smallright=301
\mathchardef\les=316
\mathchardef\gre=318
\mathchardef\leq=532
\mathchardef\grq=533
\catcode`\@=11 % @ signs are now treated as letters
\newcounter{pict@width}
\newcounter{pict@height}
\newlength{\pict@scale}
\newcommand{\psfigadd}[4]{%
\setcounter{pict@width}{1*\ratio{#2+\pict@scale/2}{\pict@scale}}
\setcounter{pict@height}{1*\ratio{#3+\pict@scale/2}{\pict@scale}}
\setlength{\unitlength}{\pict@scale}
\hbox to #2{\hspace{-\fill}\begin{picture}(\thepict@width,\thepict@height)
\put(0,0){\psfig{figure=#1,width=#2,height=#3,clip=}}
\SetScale{0.283466457}
\SetWidth{1.763889}
{#4}
\end{picture}}
}
\newcounter{pict@widthfst}
\newcounter{pict@widthscd}
\newcounter{pict@widthtot}
\newcommand{\psfigaddtwo}[7]{%
\setcounter{pict@widthfst}{1*\ratio{#2+\pict@scale/2}{\pict@scale}}
\setcounter{pict@widthscd}{1*\ratio{#2+#4+\pict@scale/2}{\pict@scale}}
\setcounter{pict@widthtot}{1*\ratio{#2+#4+#6+\pict@scale/2}{\pict@scale}}
\setcounter{pict@height}{1*\ratio{#3+\pict@scale/2}{\pict@scale}}
\setlength{\unitlength}{\pict@scale}
\hbox{\hspace{-\fill}\begin{picture}(\thepict@widthtot,\thepict@height)
\put(0,0){\psfig{figure=#1,width=#2,height=#3,clip=}}
\put(\thepict@widthscd,0){\psfig{figure=#5,width=#6,height=#3,clip=}}
\SetScale{0.283466457}
\SetWidth{1.763889}
{#7}
\end{picture}}
}
\newcommand{\psfigror}[4]{%
\setcounter{pict@width}{1*\ratio{#2+\pict@scale/2}{\pict@scale}}
\setcounter{pict@height}{1*\ratio{#3+\pict@scale/2}{\pict@scale}}
\setlength{\unitlength}{\pict@scale}
\hbox{\begin{picture}(\thepict@width,\thepict@height)
\put(0,\thepict@height){\psfig{figure=#1,width=#3,height=#2,clip=,angle=270}}
\SetScale{0.283466457}
\SetWidth{1.763889}
{#4}
\end{picture}}
}
\newcommand{\psfigrol}[4]{%
\setcounter{pict@width}{1*\ratio{#2+\pict@scale/2}{\pict@scale}}
\setcounter{pict@height}{1*\ratio{#3+\pict@scale/2}{\pict@scale}}
\setlength{\unitlength}{\pict@scale}
\hbox{\begin{picture}(\thepict@width,\thepict@height)
\put(0,0){\psfig{figure=#1,width=#3,height=#2,clip=,angle=90}}
\SetScale{0.283466457}
\SetWidth{1.763889}
{#4}
\end{picture}}
}
\catcode`\@=12 % @ signs are no longer letters
\newlength\listtextwidth

\catcode`\@=11 % @ signs are now treated as letters
\newlength{\@tabfninsert}
\newlength{\@tabfnwidth}
\newcommand{\tabfootnote}[2]{%
  \setlength{\@tabfninsert}{0.8em}
  \setlength{\@tabfnwidth}{\textwidth}
  \addtolength{\@tabfnwidth}{-\@tabfninsert}
  \addtolength{\@tabfnwidth}{-0.4em}
  \noindent\makebox[\@tabfninsert][r]{\footnotesize$^{#1}$\hfil}\hfill%
  \parbox[t]{\@tabfnwidth}{\footnotesize #2\hfill}}
\catcode`\@=12 % @ signs are no longer letters

%% file: def.tex
\newcommand{\sTop}{\protect\ensuremath{\tilde{t}}}

\newcommand{\stone}{\ensuremath{\tilde{t}_1}}

\newcommand{\rpv}{\slash\hspace{-2.5mm}{R}_{p}}
\newcommand{\rpvdue}{\slash\hspace{-1.8mm}{R}_{p}}

\newcommand{\lprim}{\protect\ensuremath{\lambda'_{131}}}

%% file: DESY-06-144.bbl
\begin{mcbibliography}{10}

\bibitem{SUSY}
Y.A.~Golfand and E.P.~Likhttman, JETP Lett. {\bf 13}, 323 (1971);\\
D.V.~Volkov and V.P.~Akulov, JETP Lett. {\bf 16}, 438 (1972);\\
D.V.~Volkov and V.P.~Akulov, Phys. Lett. {\bf B~46}, 109 (1973);\\
J.~Wess and B.~Zumino, Nucl. Phys. {\bf B~70}, 39 (1974)\relax
\bibitem{H1:SUSY}
\relax
H1 Coll., A.~Aktas \etal, Eur. Phys. J. {\bf C~36}, 425 (2004);\\
H1 Coll., A.~Aktas \etal, Phys. Lett. {\bf B~599}, 159 (2004)\relax
\relax
\bibitem{opal:stop}
OPAL Coll., G.~Abbiendi \etal, Eur. Phys. J. {\bf C~33}, 149 (2004)\relax
\relax
\bibitem{aleph:stop}
ALEPH Coll., A.~Heister \etal, Eur. Phys. J. {\bf C~31}, 1 (2003)\relax
\relax
\bibitem{l3:stop}
L3 Coll., P.~Achard \etal, Phys. Lett. {\bf B~524}, 65 (2002)\relax
\relax
\bibitem{cdf:stop}
CDF Coll., D.~Acosta \etal, Phys. Rev. Lett. {\bf 92}, 051803 (2004)\relax
\relax
\bibitem{butt}
%  J.~Butterworth and  H.~Dreiner, {\it R-Parity Violation at HERA}, Nucl. Phys. {\bf B~397}, 3 (1993)\relax
J.~Butterworth and  H.~Dreiner, Nucl. Phys. {\bf B~397}, 3 (1993)\relax
\bibitem{isr}
C.F.~Weizs\"acker, Z. Phys. {\bf 88}, 612 (1934);\\
E.J.~Williams, Phys. Rev. {\bf 45}, 729 (1934)\relax
\bibitem{nlo}
%T.~Plehn, H.~Spiesberger, M.~Spira, P.M.~Zerwas, Z. Phys. {\bf C~74}, 611 (1997)\relax
T.~Plehn \etal, Z. Phys. {\bf C~74}, 611 (1997)\relax
\relax
\bibitem{APV}
H.K.~Dreiner, {\it Perspectives on Supersymmetry}, Ed. by G.L. Kane, World Scientific, (1997), hep-ph/9707435\relax
\relax
\bibitem{epj:c19:397}
L3 Coll., M.~Acciarri \etal, Eur. Phys. J. {\bf C~19}, 397 (2001)\relax
\relax
\bibitem{polonsky}
%N.~Polonsky, {\it Supersymmetry: Structure and Phenomena}, hep-ph/0108236\relax
%N.~Polonsky, Lect. Notes Phys. {\bf M~68}, 1 (2001)\relax
H.P.~Nilles, Phys. Rept. {\bf 110}, 1 (1984);\\
H.E.~Haber and G.L. Kane, Phys. Rept. {\bf 117}, 75 (1985)\relax
\relax
\bibitem{msugra}
R.~Barbieri, S.~Ferrara and C.A.~Savoy, Phys. Lett. {\bf B~119}, 343 (1982);\\
L.J.~Hall, J.~Lykken and S.~Weinberg, Phys. Rev. {\bf D~27}, 2359 (1983);\\
P.~Nath, R.~Arnowitt and A.H.~Chamseddine, Nucl. Phys. {\bf B~227}, 121 (1983)\relax
\bibitem{zeus:1993:bluebook}
ZEUS Coll., U.~Holm \etal, {\it The ZEUS Detector}. Status Report (unpublished), DESY (1993), available on {\tt http://www-zeus.desy.de/bluebook/bluebook.html}\relax
% \relax
%ctd
\bibitem{nim:a279:290}
 N.~Harnew \etal, Nucl. Inst. Meth. {\bf A~277}, 176 (1989);\\
 B.~Foster \etal, Nucl. Inst. Phys. Proc. Suppl. {\bf B~32}, 181 (1993);\\
 B.~Foster \etal, Nucl. Inst. Meth. {\bf A~338}, 254 (1994)\relax
%cal
\bibitem{nim:a309:77}
M.~Derrick \etal, Nucl. Inst. Meth.  {\bf A~309}, 77 (1991);\\
A.~Andresen \etal, Nucl. Inst. Meth.  {\bf A~309}, 101 (1991);\\
A.~Caldwell \etal, Nucl. Inst. Meth.  {\bf A~321}, 356 (1992);\\
A.~Bernstein \etal, Nucl. Inst. Meth.  {\bf A~336}, 23 (1993)\relax
%lumi
\bibitem{Desy-92-066}
J.~Andruszk\'ow \etal, Preprint DESY-92-066, DESY, 1992;\\
ZEUS Coll., M.~Derrick \etal, Z. Phys. {\bf C~63}, 391 (1994);\\
J.~Andruszk\'ow \etal, Acta Phys. Pol. {\bf B~32}, 2025 (2001)\relax
\bibitem{susygen} N.~Ghodbane \etal, {\it SUSYGEN 3,} hep-ph/9909499\relax
\relax
\bibitem{pythia} T.~Sj\"ostrand \etal, Comp. Phys. Comm. {\bf 135}, 238 (2001)\relax
\bibitem{suspect}
A.~Djouadi, J.~Kneur and G.~Moultaka, {\it SuSpect: a Fortran Code for the Supersymmetric and Higgs Particle Spectrum in the MSSM}, hep-ph/0211331\relax
\relax
\bibitem{heracles}
  A.~Kwiatkowski, H.~Spiesberger and H.J.~Mohring, Comp. Phys. Comm. {\bf 69}, 155 (1992)\relax
\relax
\bibitem{django}
K.~Charchula, G.A.~Schuler  and  H.~Spiesberger,  Comp. Phys. Comm. {\bf 81}, 381 (1994)\relax
\relax
\bibitem{cteq5}
  CTEQ Coll., H.L.~Lai et al., Eur. Phys. J. {\bf C~12}, 375 (2000)\relax
\relax
\bibitem{ariadne}
  L.~L\"onnblad, Comp. Phys. Comm. {\bf 71}, 15 (1992)\relax
\relax
\bibitem{lepto} 
  G.~Ingelman, A.~Edin  and J.~Rathsman,   Comp.   Phys.   Comm.   {\bf 101},   108   (1997)\relax
\relax
\bibitem{jetset}
  T.~Sj\"ostrand, Comp. Phys. Comm. {\bf 39}, 347 (1986);\\
  T.~Sj\"ostrand and M.~Bengtsson, Comp. Phys. Comm. {\bf 43}, 367 (1987);\\
  T.~Sj\"ostrand, Comp. Phys. Comm. {\bf 82}, 74 (1994)\relax
\relax
\bibitem{herwig}
  G.~Marchesini \etal, Comp. Phys. Comm. {\bf 76}, 465 (1992)\relax
\relax
\bibitem{cteq4}
  CTEQ Coll., H.L.~Lai et al., Phys. Rev. {\bf D~55}, 1280 (1997)\relax
\relax
\bibitem{grvg}
  M.~Gluck et al., Phys. Rev. {\bf D~46}, 1973 (1992)\relax
\relax
%\bibitem{cteq7}
%J.~Pumplin, D.~R.~Stump, J.~Huston, H.~L.~Lai, P.~Nadolsky and W.~K.~Tung, JHEP {\bf 0207}, 012 (2002)\relax
%J.~Pumplin \etal, JHEP {\bf 0207}, 012 (2002)\relax
%\relax
%\bibitem{pythia}
%  T.~Sj\"ostrand, Comp. Phys. Comm. {\bf 82}, 74 (1994)\relax
%\relax
\bibitem{ZEUS:LQ}
ZEUS Coll., S.~Chekanov \etal, Phys. Rev. {\bf D~68}, 052004 (2003)\relax
\relax
\bibitem{ZEUS:NC}
ZEUS Coll., S.~Chekanov \etal, Phys. Rev. {\bf D~70}, 052001 (2004)\relax 
\relax
\bibitem{ZEUS:CC}
ZEUS Coll., S.~Chekanov \etal, Eur. Phys. J. {\bf C~32}, 16 (2003)\relax
\relax
\bibitem{thesis}
A.~Montanari, Ph.D. thesis, Alma Mater Universit\`a di Bologna (2001), (unpublished)\relax
\relax
\bibitem{EM}
ZEUS Coll., J.~Breitweg \etal, Z. Phys. {\bf C~74}, 207 (1997)\relax
\relax
\bibitem{DA}
S.~Bentvelsen, J.~Engelen and P.~Kooijman, {\it Proc.
of the Workshop on Physics at HERA}, W.~Buchm\"uller and G.~Ingelman (eds.), Vol~1, p.23. Hamburg, Germany, DESY (1992)\relax
\relax
\bibitem{JB}
F.~Jacques and A.~Blondel,  {\it Proceedings of the Study
    for an ep Facility for Europe,} U.~Amaldi (ed.), p.~391. Hamburg,
  Germany (1979)\relax
\relax
\bibitem{cteqsys}
J.~Pumplin \etal, JHEP {\bf 0207}, 012 (2002)\relax
\relax
%\bibitem{LEPTO}
%G.~Ingelman, A.~Edin  and J.~Rathsman, Comp. Phys. Comm. {\bf 101}, 108 (1997)\relax
%\relax
%\bibitem{cteq6}
%  CTEQ Coll., H.L.~Lai \etal, Phys. Rev. {\bf D~55}, 1280 (1997)\relax
%\relax
\bibitem{apv} 
P.~Langacker, Phys. Lett. {\bf B~256}, 277 (1991)\relax
\relax
\bibitem{D0:lq}
D0 Coll., V.M.~Abazov \etal, Phys. Rev. D Rapid Comm. {\bf 71}, 071104(R) (2005)\relax
\relax
\bibitem{cgm}
S.~Chakrabarti, M.~Guchait, N.K.~Mondal, Phys. Rev. {\bf D~68}, 015005 (2003)\relax
\relax
\bibitem{lorenzo}
L.~Bellagamba, hep-ex/0611012\relax
\relax
\end{mcbibliography}
